\documentclass[twocolumn,epjc3,numbook]{svjour3}
\RequirePackage[T1]{fontenc}
\RequirePackage{graphicx}
\RequirePackage{flushend}
\RequirePackage[numbers,sort&compress]{natbib}
\RequirePackage[colorlinks,citecolor=blue,urlcolor=blue,linkcolor=blue]{hyperref}
\RequirePackage{newtxtext,newtxmath}
\let\vec\mathbf 
\counterwithin{figure}{section}
\numberwithin{equation}{section}
\RequirePackage{cleveref}
\crefname{equation}{}{} 
\crefname{table}{Table}{Tables} 
\crefname{figure}{Figure}{Figures} 
\journalname{Eur. Phys. J. C}
\begin{document}

\title{Van der Meer Scan Luminosity Measurement and Beam--Beam Correction}

\author{Vladislav Balagura\thanksref{e1,addr1}}
\thankstext{e1}{e-mail: balagura@cern.ch}
\institute{
  \label{addr1}
  Laboratoire Leprince-Ringuet, CNRS/IN2P3, \'Ecole polytechnique, Institut Polytechnique de Paris
}


\date{Received: date / Accepted: date}

\maketitle

\begin{abstract}
  The main method for calibrating the luminosity at Large Hadron Collider
  (LHC) is van der Meer scan where the beams are swept transversely across
  each other. This beautiful method was invented in 1968. Despite the
  honourable age, it remains the preferable tool at hadron colliders. It
  delivers the lowest calibration systematics, which still often dominates the
  overall luminosity uncertainty at LHC experiments. Various details of the
  method are discussed in the paper. One of the main factors limiting
  proton--proton van der Meer scan accuracy is the beam--beam electromagnetic
  interaction.  It modifies the shapes of the colliding bunches and biases the
  measured luminosity.  In the first years of operation, four main LHC
  experiments did not attempt to correct the bias because of its
  complexity. In 2012 a correction method was proposed and then subsequently
  used by all experiments. It was based, however, on a simplified linear
  approximation of the beam--beam force and, therefore, had limited
  accuracy. In this paper, a new simulation is presented, which takes into
  account the exact non-linear force.  Depending on the beam parameters, the
  results of the new and old methods differ by $\sim1\%$. This needs to be
  propagated to all LHC cross-section measurements after 2012. The new
  simulation is going to be used at LHC in future luminosity calibrations.
\end{abstract}

\section{Van der Meer scan} 
\label{sec:vdm}
\subsection{Calibration methods}
\label{sec:method}

An absolute value of the luminosity or the cross section can be measured at an
accelerator by separating the beams in the transverse plane and performing the
so-called {\it van der Meer scan}~\cite{vanderMeer:296752}. To illustrate the
idea, let us consider the collision of two bunches with $N_{1,2}$ particles
moving in the opposite directions. If the first bunch is separated by
$-\Delta x$, $-\Delta y$ in the plane perpendicular to the beams, the average
number of interactions $\mu$ with the cross section $\sigma$ normalized by the
number of particles is
\begin{align}
  \label{eq:prevdm}
&  \frac{\mu(\Delta x,\, \Delta y)}{N_1N_2} = \sigma \frac{L}{N_1N_2} \nonumber\\
&= \sigma \iint \rho_1(x_2 + \Delta x,\, y_2 + \Delta y) \rho_2(x_2,y_2) dx_2\, dy_2,
\end{align}
where the subscript ``2'' of the coordinates $x_2$, $y_2$ refers to the
stationary second beam, $L$ is the integrated luminosity and
$\rho_{1,2}(x_2,y_2)$ are the normalized transverse particle densities of the
unseparated bunches when $\Delta x = \Delta y = 0$.  For example, if
$\rho_1(x,y)$ is a delta-function $\delta(x,y)$, the shifted density
$\delta(x+\Delta x, y+\Delta y)$ peaks at $-\Delta x,-\Delta y$ (note the
minus sign).  Integration over $\Delta x$ and $\Delta y$ drastically
simplifies \cref{eq:prevdm} since
\begin{equation}
\label{eq:5}
 \int \rho_1(x_2 + \Delta x,\, y_2 + \Delta y) \rho_2(x_2,y_2) dx_2\, dy_2\, d\Delta x\, d\Delta y = 1,
\end{equation}
as can be easily proved by substituting $x_1=x_2+\Delta x,\ y_1=y_2+\Delta y$.
In the new variables the integrals $\iint \rho_1(x_1,\, y_1)dx_1\,dy_1$ and
$\iint \rho_2(x_2,\, y_2)dx_2\,dy_2$ decouple and reduce to unity by
definition.

From \cref{eq:prevdm} and \cref{eq:5} we obtain van der Meer formula
\begin{equation}
  \label{eq:vdm}
 \sigma=  \iint \frac{\mu(\Delta x,\, \Delta y)}{N_1N_2} d\Delta x\, d\Delta y.
\end{equation}
The ratio 
\begin{equation}
\label{eq:6}
\mu_{sp}=\mu/N_1N_2.
\end{equation}
is often called the ``specific'' number of interactions.  Their total number
accumulated during the scan, $\iint\mu_{sp}d\Delta x\, d \Delta y$,
gives $\sigma$.

Though van der Meer method is well known, the formula \cref{eq:vdm} is
sometimes reduced in the literature to the Gaussian bunch densities or the
case when $\mu_{sp}$ is distributed independently in $x$ and $y$, discussed in
the next section.  Here, we present the method in its full generality. This is
required, in particular, for explaining the novel two-dimensional
scans~\cite{twoD}, which will probably be in wide use at LHC in Run 3.  After
analyzing van der Meer scans for several years, the author tries to share the
accumulated experience on various calibration details in the first half of the
paper. The discussion is concentrated on the general method and its accuracy
but not on the detector effects varying from one luminometer to the other. The
second half of the paper is fully devoted to the beam--beam electromagnetic
interaction, which is one of the main factors limiting the accuracy.

The derivation of \cref{eq:vdm} uses only $\mu_{sp}$ but not $\mu$ or
$N_{1,2}$ separately.  Therefore, it remains valid even if $\mu$ and $N_1N_2$
change arbitrarily but proportionally during the scan, eg. due to a gradual
decrease of beam currents with time. It is required, however, that
$\rho_{1,2}$ {\it remain constant}.

Equation \cref{eq:vdm} can also be understood from other but equivalent
perspective.  The transverse movements of the first bunch {\it smear}
and``wash out'' its profile $\rho_1$, so that effectively it becomes {\it
  constant} $\bar{\rho_1}$. This reduces the complicated overlap integral to
$\iint\bar{\rho_1}\rho_2 dx_2\,dy_2=\bar{\rho_1}$, ie. to unity
$\bar{\rho_1}=\int\rho_1d\Delta x\,\Delta y=1$ if $\rho_{1,2}$ are normalized.
Equivalently, the scan can be viewed from the transverse position of the first
beam ie. in the coordinates $x_1,\ y_1$. Then the first beam is stationary
while the second moves. In this case $\rho_2(x_1-\Delta x,y_1-\Delta y)$ is
``washed out'' by $\Delta x,\,\Delta y$ movements, and the overlap integral
reduces to $\int\rho_1dx_1\,dy_1=1$ and drops out as before.

To make a parallel translation of one beam in one transverse direction, one
needs 4 magnets placed at the corners of the trapezoid-like beam trajectory.
Therefore, to steer two beams in two directions one needs
$4\times2\times2 = 16$ magnets per one interaction point. It is not easy to
synchronize precisely all of them and ensure a parallel translation of the
beams with constant speeds.  Therefore, the scan, eg. at LHC, is performed not
``dynamically'' using the single continuous pass, but stepwise.  Ie. the
function $\mu_{sp}(\Delta x,\, \Delta y)$ is measured only in the predefined
set of discrete points and interpolated between them or fitted to some
analytic function to get the final integral. In moving the beams from one
point to the next, one waits when the slowest magnet reaches it is desired
current value, and only then the luminosity measurement starts.

This beautiful method was invented by van der Meer more than 50 years ago for
the ISR accelerator~\cite{Carboni:156499}. It was proposed for
Sp$\bar{\mathrm{p}}$S~\cite{SppS}, successfully applied with various
modifications at RHIC~\cite{Drees:1307872,Drees2013AnalysisOV} and
LHC~\cite{White:2010zzc,lhcb12,CMS-PAS-LUM-13-001,Aad:2013ucp,lhcb14,
  Abelev:2014epa,Aaboud:2016hhf,alice:2160174,cms15,cms16,cms17,atlas19}. The
method delivered a record accuracy between 0.7\% and a few percent.

At $e^+$-$e^-$ colliders van der Meer method is biased by strong beam--beam
interactions. The luminosity is usually measured ``indirectly'' by counting
Bhabha scattering $e^+e^-\to e^+e^-$ events, which have a high cross-section
precisely known from quantum electrodynamics.  Contrary to that, in the
collisions of non-elementary hadrons it is difficult to find a physical
process with accurately predicted cross-section and convenient for the
detection. In the hadron accelerators, the best accuracy is achieved by
measuring the luminosity ``directly'' using its definition \cref{eq:prevdm}.

At LHC, which will only be discussed in the following, in addition to van der
Meer scans there are two alternative direct methods of the luminosity
calibration. They utilize precise vertex detectors. The first is the so-called
beam--gas imaging~\cite{Ferro-Luzzi:844569}. Here, the profiles
$\rho_{1,2}(x,y)$ of the bunches are ``revealed'' in their interactions with a
tiny amount of gas in the beam pipe. One effectively records the bunch
``photos'' using the vertex detector as a ``camera'' and the gas as a
``film''.  Unfolding the images with the vertex resolution yields $\rho_{1,2}$
densities.  To improve the accuracy, one also uses the high statistics profile
of the ``luminous region'' formed by the interactions of two bunches. This
provides a powerful constraint on the product $\rho_1\rho_2$.
The overlap integral is then calculated analytically from the
reconstructed $\rho_{1,2}$ densities.  This method is used up to now only at
LHCb~\cite{lhcb12,lhcb14}, which has a dedicated gas injection system~\cite{SMOG},
an excellent vertex detector and a flexible trigger suitable for recording
beam--gas interactions.

The second method is called the beam--beam imaging~\cite{Balagura:2011yw}. It
is very similar, but the role of the gas plays another beam effectively
``smeared'' by the transverse movements in van der Meer scan. After sweeping
eg. the first bunch, it effectively becomes a wide and uniform ``film''
independently of the initial $\rho_1$ distribution. It allows making a
``photo'' of $\rho_2$ with high statistics. Alternatively, the same vertex
distribution data can be viewed from the transverse position of the first
bunch, where it is effectively stationary.  The accumulated image gives a
``photo'' of $\rho_1$ ``filmed'' by smeared $\rho_2$.  In the original van der
Meer approach the smearing allows integrating over $\Delta x\,\Delta y$ and
reducing the overlap integral $\int\rho_1\rho_2\,dx\,dy$ to unity. In the
beam--beam imaging the same integration is applied to the luminous region
profile, ie. to the product $\rho_1\rho_2$ {\it not integrated} over $dx\,dy$,
and allows reconstructing the individual densities $\rho_{1,2}$.

Up to now, the beam--beam images were taken at LHCb~\cite{lhcb12} and CMS
experiments~\cite{Zanetti:1357856}. Since in both imaging methods the overlap
integrals are calculated from the measured vertex distributions, they have
different systematic errors and are complementary to the ``classical'' van der
Meer approach where the vertex distributions are ignored. All three methods
might achieve similar levels of accuracy. The imaging methods are more
complicated, however, because they require a deconvolution with the vertex
resolution comparable to the transverse bunch widths.  Due to the simplicity
and sufficiency of the classical van der Meer technique, it remains the main
tool of the luminosity calibrations at LHC.

\subsection{X-Y factorization}
\label{sec:xy}
It is impossible to guide particles exactly parallel to a beam axis.
Therefore, in any accelerator the optic elements are designed such that the
particles going away from the axis are sent back and in the end ``oscillate''
in the transverse plane. This creates the transverse bunch widths and
determines the density profiles $\rho_{1,2}$. In more detail this will be
discussed in Sec.~\ref{accdynam}. To ensure stable operation, the accelerator
is designed such that the oscillatory motions are separately stable in the
transverse coordinates $x$ and $y$ and are almost independent of each
other. Any ``coupling'' between the coordinates could create extra resonances
in the oscillatory motions and, therefore, should be avoided.  The bunch
densities can often be factorized into $x$- and $y$-dependent parts:
\begin{equation}
  \label{eq:factxy}
\rho_{1,2}(x, y) = \rho_{1,2}^x(x) \cdot \rho_{1,2}^y(y).
\end{equation}
From \cref{eq:prevdm} it follows that the specific number of
interactions then also factorizes,
$\mu_{sp}(\Delta x, \Delta y) = \mu^x_{sp}(\Delta x)\cdot \mu^y_{sp}(\Delta
y)$. This is sufficient to simplify the two-dimensional integral
$\iint \mu_{sp}(\Delta x, \Delta y) d \Delta x\, d\Delta y$ and to reduce it
to a product of {\it one-dimensional integrals} along the lines
$\Delta x=\Delta x_0$ and $\Delta y=\Delta y_0$. Indeed,
\begin{align}
\sigma &= \iint \mu_{sp}(\Delta x, \Delta y) d\Delta x\, d\Delta y \nonumber\\
&=\int \mu^x_{sp}(\Delta x) d\Delta x \int  \mu^y_{sp}(\Delta y) d\Delta y\,
\frac{\mu^y_{sp}(\Delta y_0)\mu^x_{sp}(\Delta x_0)}{\mu^y_{sp}(\Delta y_0)\mu^x_{sp}(\Delta x_0)} \nonumber\\
&  = \frac{\int \mu_{sp}(\Delta x, \Delta y_0)\, d\Delta x
 \times \int \mu_{sp}(\Delta x_0, \Delta y)\, d\Delta y}
{\mu_{sp}(\Delta x_0,\Delta y_0)}.
\label{eq:fact}
\end{align}
The integrals in the enumerator can be measured in two one-dimensional scans
over $\Delta x$ at fixed $\Delta y_0$ and vice versa.  Note that the formula
is valid for any point $(\Delta x_0,\Delta y_0)$. This is rarely stressed in
the literature.  It might be advantageous to choose $(\Delta x_0,\Delta y_0)$
not far from the point of maximal luminosity to collect sufficient statistics
of interactions. There might be another advantage if the beam coordinates are
not accurately measured. The potential slow drifts of the beam orbits from
their nominal positions might affect both scanned and not scanned coordinates
and bias the luminosity measurement. The bias from not scanned coordinate is
minimized at the maximum of $\mu$ where the derivative of eg.
$\mu_{sp}(\Delta x_0, \Delta y)$ on $\Delta x_0$ is zero.

Performing a pair of one-dimensional scans instead of an expensive
two-dimensional scan allows saving the beam time. Reducing the time also
helps to minimize the influence of the slow drifts of the beam orbits if they
are not accurately measured. Therefore, at LHC the cross-sections are usually
calibrated using \cref{eq:fact} instead of \cref{eq:vdm}. This approach,
however, relies on the $x$-$y$ factorizability of $\mu$, which is good at LHC
but not perfect.  It can be violated by many factors, essentially, by any
imperfection in the accelerator leading to an $x$-$y$ coupling.

The remaining non-factorizability is usually studied using the distributions
of the interaction vertices. After unfolding with the vertex resolution they
give the products of two bunch densities.  Ideally, the shape of their
projections to one coordinate should remain invariant when scanning another
coordinate. The deviations are interpreted as the non-factorizability and are
propagated to the cross-section corrections. This procedure is complicated
because it requires the characterization of two unknown bunch shapes using
only one luminous region profile.  One can use the imaging methods to measure
directly $\rho_{1,2}$ densities~\cite{bbicms}. However, the beam--gas
interactions have limited statistics while the beam--beam imaging suffers from
the uncertainties in the beam positions and the beam--beam systematics
discussed later.  Because of the complexity of the $x$-$y$ non-factorization
studies, usually the bunch shapes are fitted assuming some smooth bunch shape
model. This makes them model-dependent.  The cross-section corrections due to
$x$-$y$ non-factorizability and the associated systematic errors are typically
at the level $\lesssim 1\%$.

The accuracy can be improved further by performing {\it two-dimensional scans
  over the central region} giving the dominant contribution to the integral in
\cref{eq:vdm}. This approach was pioneered at LHCb in
2017~\cite{twoD}. Scanning only the central region was relatively fast but
allowed evaluating $\sim90\%$ of the integral. The method is model-independent
and allows reducing the non-factorization systematics by an order of
magnitude.

\subsection{Crossing angle between the beams}
In van der Meer scans at LHC the beams are not always opposite. They may
collide at a small angle of the order $O(100~\mu rad)$. This separates the
beams outside the interaction region and suppresses possible parasitic
collisions between nominally not colliding bunches. It is not immediately
obvious how van der Meer formalism should be extended to the case of not
parallel beams. In addition, the particles in the two beams can be
different. Up to now, LHC has performed van der Meer scans with the proton and
lead ion beams. It might be not clear whether in the asymmetric proton - ion
collisions the cross-section can be calibrated in the laboratory frame, or it
is necessary to make a transformation to the center-of-mass system.

These questions were answered in~\cite{Balagura:2011yw} using two alternative
derivations.  In the first, the simple two-dimensional integral in
\cref{eq:prevdm} was extended to four dimensions and taken over
$\Delta x$, $\Delta y$ in the same way as \cref{eq:vdm} was obtained from
\cref{eq:prevdm}.  In the second, equivalent derivation the direct
mathematical calculation was substituted by simple physical arguments. They
will be elaborated in more detail below.

\begin{figure}[htbp]
\begin{center}
  \includegraphics[width=0.45\textwidth,angle=0]{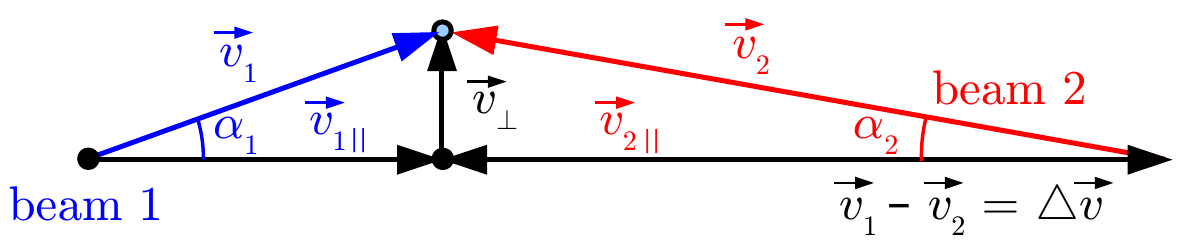}
  \vspace{-2mm}
  \caption{Decomposition of the beam particle velocities $\vec{v}_{1,2}$ to the
    perpendicular $\vec{v}_{\perp}$ and the parallel $\vec{v}_{1 \parallel}$,
    $\vec{v}_{2 \parallel}$ components with respect to their difference
    $\vec{v}_1-\vec{v}_2=\Delta \vec{v}$.}
  \label{fig1}
\end{center}
\end{figure}

Let's denote the velocities of the beam particles by $\vec{v}_{1,2}$ as shown
in Fig.~\ref{fig1}.  They can be decomposed into two parallel
($\vec{v}_{1,2}$) and one common perpendicular component $\vec{v}_{\perp}$
with respect to {\it their difference} $\vec{v}_1-\vec{v}_2 = \Delta \vec{v}$.

There exist infinitely many relativistic frames where the beams are
parallel. In any of them, \cref{eq:vdm} is valid, for example, in the
center-of-mass or the rest frame of one of the particles. It is even valid in
the frame where the particles move in the same direction one running after the
other. All such frames can be obtained by boosting the laboratory frame first
with the velocity $\vec{v}_{\perp}$ and then with an arbitrary velocity
parallel to $\Delta \vec{v}$. Indeed, in the laboratory frame boosted by
$\vec{v}_{\perp}$, or, equivalently, after the ``active'' boost of the beam
particles by $-\vec{v}_{\perp}$, their perpendicular momentum transforms to
$\vec{p}_{\perp}' = \gamma_{\perp}\vec{p}_{\perp} -
\gamma_{\perp}\vec{\beta}_{\perp}E/c$, which is zero since by construction
$\vec{\beta}_{\perp} = \vec{p}_{\perp}c/E$. Here, $c$ is the speed of light,
$E$ is the energy in the laboratory system,
$\vec{\beta}_{\perp}=\vec{v}_{\perp}/c$ and
$\gamma_{\perp}=(1-\beta_{\perp}^2)^{-1/2}$ are the beta- and gamma-factors of
the boost, respectively.  Therefore, after the active boost by
$-\vec{v}_{\perp}$ the beams become parallel to $\Delta \vec{v}$ and any
further boost along $\Delta\vec{v}$ preserves this parallelism.

Note that although the perpendicular velocity after the boost by
$-\vec{v}_{\perp}$ becomes a simple difference
$\vec{v}_{\perp}-\vec{v}_{\perp} = \vec{0}$, the remaining velocity
$\vec{v}_{\parallel}'$ is, of course, not equal to the difference
$\vec{v} - \vec{v}_{\perp} = \vec{v}_{\parallel}$. This would be the case for
the Galilean but not for the Lorentz transformation.
The correct formula follows most easily from the {\it opposite} transformation
from the primed to the laboratory frame:
$\vec{p}_{\parallel}=\vec{p}_{\parallel}'$,
$E=\gamma_{\perp}E'$, so
$\vec{v}_{\parallel}' =\gamma_{\perp}\vec{v}_{\parallel}$.

Let's find out how van der Meer formula \cref{eq:vdm}, proved in {\it the
  primed coordinates} with the parallel beams, modifies in the laboratory
frame.  The quantities $\mu$, $N_{1,2}$ and $\sigma$ are Lorentz-invariant,
only the transverse area $d\Delta x\,d\Delta y$ is not. Following notations of
this subsection, the latter will be denoted in {\it the primed coordinates} as
$d\Delta x'\,d\Delta y'$. Its transformation to the laboratory system requires
some explanations given below. This material complements the discussion
in~\cite{Balagura:2011yw}.

The space-time coordinates $x=(t,\,\vec{x})$ of any particle moving with the
four-momentum $p=(E,\,\vec{p})$ satisfy the equation
\begin{equation}
\label{eq:10}
x-x^0 = \lambda p,
\end{equation}
where $\lambda$ is a free parameter and $x^0=(t^0,\,\vec{x}^0)$ is an
arbitrary point on the particle trajectory corresponding to $\lambda=0$.
Four equations in \cref{eq:10} with one free parameter define a line in
the four-dimensional space.  The values of $\lambda$ uniquely label the line
points and can be expressed, for example, via the time coordinate:
$\lambda = (t-t^0)/E$.

The beam displacements $\Delta = (0, \vec{\Delta})$ during van der Meer scan
change the $x^0$ parameter:
\begin{equation}
\label{eq:43}
x-x^0-\Delta = \lambda p.
\end{equation}
Note that the definition of the scan implies that the beams are moved only
spatially, so there is {\it no time component} in $\Delta$.

In the following, it will be convenient to decompose spatial vectors into
three components: parallel to $\vec{v}_{\perp}$, to the vector-product
$[\Delta \vec{v}\times\vec{v}_{\perp}]$ and to $\Delta \vec{v}$.  They will be
denoted by the subscripts $\perp$, $\parallel\times\perp$ and $\parallel$,
respectively, eg.
\begin{equation}
\label{eq:99}
\Delta =
(0,\,\Delta_{\perp},\Delta_{\parallel\times\perp},\Delta_{\parallel}).
\end{equation}
The $\parallel\times\perp$ component is perpendicular to the plane of
Fig.~\ref{fig1}.  Ideally, the beam displacements should be orthogonal to
$\Delta \vec{v}$ but we consider below the general case
$\Delta_{\parallel}\ne0$.

After the active boost by $-\Delta\vec{v}$, the line defined by \cref{eq:43}
transforms to
\begin{equation}
\label{eq:44}
x'-x^{0\prime}-\Delta' = \lambda p'.
\end{equation}
Note that the solutions $x$, $x'$ of \cref{eq:43,eq:44} for the
same $\lambda$ correspond to the same four-dimensional point in the laboratory
and primed frames, respectively.

After the boost, the beam displacement
\begin{equation}
\label{eq:45}
\Delta' =
(-\beta_{\perp}\gamma_{\perp}\Delta_{\perp},\,\,
\gamma_{\perp}\Delta_{\perp},\,
\Delta_{\parallel\times\perp},\,
\Delta_{\parallel})
\end{equation}
acquires the time component
$\Delta_t'=-\beta_{\perp}\gamma_{\perp}\Delta_{\perp}$. In other words, it is
impossible to transform a spatial scan in the laboratory to merely a spatial
scan in the primed frame. To resolve this complication one can use the
following argument.  Any system with the {\it parallel} beams is special in
the sense that the number of interactions created by two particles is
determined only by the transverse distance between their lines. It does not
depend on the initial positions of the particles
$x^{0\prime}_{\parallel} + \Delta_{\parallel}'$ along the lines or on the
initial time $t^{0\prime}+\Delta_t'$ since the particles are anyway assumed to
travel from $t'=-\infty$ to $t'=+\infty$. Contrary to the transverse initial
coordinates, the time and longitudinal shifts do not change the particle
line. Therefore, the scan with $\Delta'$ from \cref{eq:45} is equivalent to
the one with
\begin{equation}
\label{eq:47}
\tilde\Delta' = (0,\, \gamma_{\perp}\Delta_{\perp},\, \Delta_{\parallel\times\perp},\,0),
\end{equation}
where the time and longitudinal coordinates are simply set to zero.
Comparing this equation with \cref{eq:99} one can see that the beam
displacements $\Delta_{\parallel\times\perp}$ transform from the laboratory to
the primed system without any changes while the displacements $\Delta_{\perp}$
are ``extended'' by the $\gamma_{\perp}$-factor. Therefore, the area element
$d\Delta x'\,d\Delta y'$ in the primed coordinates is larger than in the
laboratory system by $\gamma_{\perp}$:
\begin{equation}
\label{eq:97}
d\Delta x'\,d\Delta y' = \gamma_{\perp}d\Delta x\,d\Delta y.
\end{equation}

Note that one can invert the arguments and make the opposite boost from the
primed scan with the displacements
$ (0, \Delta'_{\perp},\, \Delta'_{\parallel\times\perp},\, 0)$ to the
laboratory system:
\[
\Delta = (\beta_{\perp}\gamma_{\perp}\Delta'_{\perp},\,\,
\gamma_{\perp}\Delta'_{\perp},\,
\Delta'_{\parallel\times\perp},\, 0).
\]
Here, the $\gamma_{\perp}$-factor again appears in the transverse components
{\it but now in the laboratory system.}  Concluding from this formula that
$\gamma_{\perp}d\Delta x'\,d\Delta y' = d\Delta x\,d\Delta y$ {\it with
  $\gamma_{\perp}$ on the opposite side} would be a mistake, however, since
the time and longitudinal components can be freely changed and zeroed {\it
  only} in the primed frame with the {\it parallel} beams. Setting
$\Delta_t=\beta_{\perp}\gamma_{\perp}\Delta'_{\perp}$ to zero in the formula
above is not allowed and spoils the equivalence of the scans in the laboratory
and the primed frames.

Using \cref{eq:97}, the cross-section formula in the laboratory system can
finally be written as
\begin{equation}
\label{eq:18}
\sigma= \gamma_{\perp}\int \frac{\mu(\Delta x, \Delta y)}{N_1N_2}d\Delta x\,d\Delta y.
\end{equation}

Note that the relativistic correction $\gamma_{\perp}$ depends on the
velocities but not on the momenta or masses of the particles. For example, it
coincides for proton and lead ion beams if their velocities are the same. The
formula is relativistically invariant and is valid in any frame. The area
element $d\Delta x\,d\Delta y$ by definition lies in the plane perpendicular
to $\Delta \vec{v}$. Let's denote this plane by $P$. If the scan plane
$\tilde P$ is inclined with respect to $P$ at an angle $\alpha$, an area
element $d\tilde\Delta x\,d\tilde\Delta y$ on $\tilde P$ should be projected
to $P$, ie.
\begin{equation}
\label{eq:100}
\sigma= \gamma_{\perp}\int \frac{\mu(\tilde\Delta x, \tilde\Delta y)}{N_1N_2}
\cos\alpha\, d\tilde\Delta x\,d\tilde\Delta y.
\end{equation}
The longitudinal translations along $\Delta \vec{v}$ do not matter.

The beam crossing angles at LHC are small, so $\beta_{\perp}<10^{-3}$ and
$\gamma_{\perp}-1<10^{-6}$. Therefore, the relativistic correction at LHC can
be safely neglected.

The luminosity, however, is modified as it can be seen from the following.
The typical longitudinal sizes of the LHC bunches $\sigma_{iL}$ along the beam
$i=1,2$ in the laboratory frame are 5--10~cm. They are much larger than the
transverse sizes $\sigma_{iT}\sim 100\, \mu m$.  Since
$\gamma_{\perp}\approx 1$, the Lorentz and Galilean transformations from the
laboratory to primed coordinates are almost equivalent. The latter preserves
the bunch shapes.  Therefore, in the primed system the longitudinal spread
$\sigma_{iL}$ gets projected to $\vec{v}_{\perp}$ at the angles $\alpha_{1,2}$
shown in Fig.~\ref{fig1}.  For the Gaussian bunches this increases the primed
transverse width $\sigma_{i\perp}'$ in the beam crossing plane to
\begin{equation}
\label{eq:27}
\sigma_{i\perp}'\approx\sigma_{iT}\sqrt{1+(\alpha_i\,\sigma_{iL}/\sigma_{iT})^2}.
\end{equation}
This formula is valid up to the second-order $\alpha_i$-terms {\it enhanced
  by} $\sigma_{iL}/\sigma_{iT}\gg 1$ factor.

Note that often in the literature $\sigma_{i\perp}'$ is expressed in the form of a
rotation as
$\sqrt{(\sigma_{iT}\cos\alpha_{i})^2 + (\sigma_{iL}\sin\alpha_{i})^2 }$.
Writing $\cos\alpha_i= 1 - \alpha_{i}^2/2+\ldots$ instead of
unity as in \cref{eq:27} implies its validity at least up to the terms
$\propto\alpha_{i}^2$ {\it not enhanced by} $\sigma_{iL}/\sigma_{iT}$.  At
this level of accuracy one can not neglect the difference between Lorentz and
Galilean transformations, however, and should take into account the
relativistic corrections.

The exact formula can be obtained using the same formalism as for van der Meer
scan above.  Let's interpret the parameter $\Delta=(0,\vec{\Delta})$ in
\cref{eq:43} not as the beam displacement but as a {\it stochastic} variable
describing the spatial spread of the particles in the bunch in the {\it
  laboratory} frame.  For example, $\vec{\Delta}$ can be a zero-mean random
variable normally and independently distributed along the transverse $x$, $y$
and longitudinal axes with the standard deviations $\sigma_{x,y,L}$,
respectively.  Let's consider the general case when neither $x$ nor $y$ lies
in the crossing plane of \cref{fig1}. Then, the transverse bunch widths in the
laboratory frame along $\perp$ and $\parallel\times\perp$ directions are
\begin{align}
\label{eq:116}
  \sigma_{i\perp}&=\sqrt{\left[\sigma_{ix}\cos(x_i,\perp)\right]^2+
                   \left[\sigma_{iy}\cos(y_i,\perp)\right]^2+\left[\sigma_{iL}\sin\alpha_i\right]^2}\nonumber\\
  \sigma_{i\parallel\times\perp}&=\sqrt{\left[\sigma_{ix}\cos(x_i,\parallel\!\times\!\perp)\right]^2+
                                  \left[\sigma_{iy}\cos(y_i,\parallel\!\times\!\perp)\right]^2}.
\end{align}
Here, the notation like $\cos(x_i,\perp)$ denote the cosine of the angle
between the $x$ direction of the $i$-th bunch and $\vec{v}_{\perp}$.
According to \cref{eq:47}, the bunch spread $\sigma_{\perp}$ is multiplied by
the $\gamma_{\perp}$-factor in the {\it primed} frame, while
$\sigma_{\parallel\times\perp}$ remains unchanged:
\begin{equation}
\label{eq:117}
\sigma_{i\perp}'=\gamma_{\perp}\sigma_{i\perp},\quad \sigma_{i\parallel\times\perp}'=\sigma_{i\parallel\times\perp}.
\end{equation}

If one of the transverse axes, eg. $x$, lies in the crossing plane, this
simplifies to
\begin{equation}
\label{eq:118}
\sigma_{i\perp}'=\gamma_{\perp}\sqrt{\left[\sigma_{ix}\cos\alpha_i\right]^2+
  \left[\sigma_{iL}\sin\alpha_i\right]^2},
\ \ \ 
\sigma_{i\parallel\times\perp}'=\sigma_y.
\end{equation}

For ultrarelativistic beams with $v_{1,2}\approx c$, like at LHC, and arbitrary
$\alpha_{1,2}$, which should be approximately equal in this case,
$\alpha_1\approx\alpha_2\approx\alpha$, the beta- and gamma-factors can be
expressed via the angle $\alpha$:
\begin{equation}
\label{eq:101}
\beta_{\perp}\approx \sin\alpha,\quad
\gamma_{\perp}\approx 1/\sqrt{1-\sin^2\alpha} = 1/\cos\alpha.
\end{equation}
This gives
\begin{equation}
\label{eq:39}
  \sigma_{i\perp}'\approx\sqrt{\sigma_{ix}^2+(\sigma_{iL}\tan\alpha)^2}.
\end{equation}
One can see that $\sigma_{ix}$ is not multiplied by $\cos\alpha$.  Instead of
the rotation formula, for $\sigma_{i\perp}'$ one should use either exact
\cref{eq:116}, \cref{eq:117}, \cref{eq:118} or one of the approximate equations \cref{eq:27},
\cref{eq:39}.

The smallness of $\alpha_i$ at LHC is partially compensated by the large
$\sigma_{iL}/\sigma_{iT}$ ratio, so due to the crossing angle the effective
transverse widths $\sigma_{i\perp}'$ increase by 5--20\%. The luminosity reduces
by the same amount. Once again, this does not modify van der Meer formula
\cref{eq:18}, since the normalization $\int\rho_{i}(x,\, y)\,dx\,dy=1$ remains
invariant. For broader bunches, one just needs to enlarge proportionally the
region of integration.


Note that if the transverse bunch densities $\rho_{1,2}$ factorize in the
primed directions $x'$ and $y'$ but none of them lies in the crossing plane of
\cref{fig1}, $\sigma_{iL}$ has non-zero projections on both $x'$ and
$y'$. This makes $x'$ and $y'$ distributions slightly correlated and to some
extent breaks the $x'$-$y'$ factorizability.

\subsection{Luminosity calibration accuracy}
Van der Meer scan is the main tool of the absolute luminosity calibration at
LHC. For the given beam particles and the LHC energy, the scan is performed in
every experiment at least once a year to check the stability of the luminosity
detectors. One or two LHC fills with carefully optimized experimental
conditions are allocated for this purpose. The accuracy is determined by
various factors discussed below. The overall calibration uncertainty is
typically 1-2\%.

The calibration constant is then propagated to the luminosity of the whole
physics sample using linear luminometers. In ATLAS and CMS operating at higher
pile-up $\mu$-values, the accurate linearity is required in larger dynamic
range since van der Meer scans are typically performed with $\mu\lesssim
1$. The uncertainties caused by the deviations from the linearity due to
irradiation ageing, long-term instabilities etc. are reduced by comparing
several luminometers with different systematics. Therefore, the overall
luminosity uncertainty is often dominated by the calibration error.

Averaging over many colliding bunch pairs and several van der Meer scans often
reduces the statistical error of the calibration to a negligible level. The
systematics from the bunch population $N_1N_2$ measurements is typically at
the level $\sim0.2-0.3\%$ except in the very first LHC scans in 2010 with
low-intensity beams.

The cross-section has the dimensionality of a length unit square. It appears
in van der Meer formula \cref{eq:18} due to the integration over $\Delta x$
and $\Delta y$. Any error in the beam displacements directly affects the
cross-section. An accurate measurement of $\Delta x$, $\Delta y$ scale is
performed in a dedicated ``length scale calibration'' (LSC), which always
accompanies van der Meer scans. The simplest LSC is described below.

Let's assume that the {\it true} beam positions $\vec{\Delta}_i$ in the
laboratory frame for the beam $i=1,2$ can be written as
\begin{equation}
\label{eq:21}
\vec{\Delta}_i = \alpha_i \vec{a}_i+ \beta_i \vec{b}_i + \vec{\Delta}_i^0,
\end{equation}
where $\vec{a}_i$, $\vec{b}_i$ are unknown vectors close but not exactly equal
to the unit $x$, $y$-vectors and $\alpha_i$, $\beta_i$ are the {\it nominally
  set} values for $x$, $y$ beam movements, respectively. The latter are known
exactly. Potential nonlinearities in the $\vec{\Delta}_i$ dependence on
$\alpha_i,\beta_i$ eg. due to beam orbit drifts, are neglected here but will be
briefly discussed later. The constant vectors $\vec{\Delta}_i^0$ corresponding
to $\alpha_i=\beta_i=0$ are also unknown.

In the simplest LSC the beams are {\it nominally} displaced by the same
amount, ie. with $\alpha_1=\alpha_2=\alpha_{LSC}$,
$\beta_1=\beta_2=\beta_{LSC}$. If the bunches have equal shapes, the center of the luminous
region $\vec{O}_{LSC}$ is positioned in the middle between them,
\begin{equation}
\label{eq:28}
\vec{O}_{LSC}=\alpha_{LSC}\,\frac{\vec{a}_1+\vec{a}_2}{2} + \beta_{LSC}\,\frac{\vec{b}_1 +\vec{b}_2}{2} +
\frac{\vec{\Delta}_1^0+\vec{\Delta}_2^0}{2}.
\end{equation}
It can be accurately measured by the vertex detectors.

Most often van der Meer scans at LHC are performed such that the beams are
displaced symmetrically in opposite directions corresponding to
$\alpha_1=-\alpha_2=2\alpha^{sym}$, $\beta_1=-\beta_2=2\beta^{sym}$. The
advantage of the symmetric scan is that it allows to reach maximal separations
in the limited allowed range of the beam movements.  The distance between the
beams $\vec{\Delta}_{12}=\vec{\Delta}_1-\vec{\Delta}_2$ is then
\begin{equation}
\label{eq:4}
\vec{\Delta}_{12}=
\alpha^{sym}\,\frac{\vec{a}_1+\vec{a}_2}{2} + \beta^{sym}\,\frac{\vec{b}_1 +\vec{b}_2}{2}+
\frac{\vec{\Delta}_1^0-\vec{\Delta}_2^0}{2}.
\end{equation}
A comparison of~\cref{eq:28} and~\cref{eq:4} shows that the measurements of
$\vec{O}_{LSC}$ in the vertex detector are sufficient to calibrate the scales
$(\vec{a}_1+\vec{a}_2)/2$, $(\vec{b}_1+\vec{b}_2)/2$ necessary for the
symmetric scans. The fact that the shapes of the colliding bunches are
different can normally be neglected at LHC after averaging over many colliding
bunch pairs.

Non-symmetric scans depend on other linear combinations than
$(\vec{a}_1+\vec{a}_2)/2$ and $(\vec{b}_1 +\vec{b}_2)/2$ measurable in
\cref{eq:28}.  The simplest way to calibrate the lengths $|\vec{a}_{1,2}|=a_{1,2}$
and $|\vec{b}_{1,2}|=b_{1,2}$ individually is to use the measurements of the
luminosity. Ideally it should be stationary during LSC. Any small variation
indicates that the distance between the beams changes.

For example, let's consider the LSC in the $x$-direction. Assuming $x$-$y$
factorizability and  neglecting the angle between $\vec{a}_{1,2}$
  and the $x$-axis, one arrives at the scalar equations
\begin{align}
\label{eq:103}
\Delta_{i,x} &= \alpha_{LSC}\, a_i + \Delta_{i,x}^0,\nonumber \\
\Delta_{12,x} &= \alpha_{LSC}(a_1-a_2) + (\Delta_{1,x}^0-\Delta_{2,x}^0),\nonumber \\
O_{LSC,x}&=\alpha_{LSC}\,\frac{a_1+a_2}{2}+\frac{\Delta_{1,x}^0+\Delta_{2,x}^0}{2}.
\end{align}
The movement of the luminous region between any two LSC points,
$O^1_{LSC,x}-O^2_{LSC,x}$, allows to measure the average $x$-scale
\begin{equation}
\label{eq:102}
\frac{a_1+a_2}{2} =  \frac{O^1_{LSC,x}-O^2_{LSC,x}}{\alpha_{LSC}^1-\alpha_{LSC}^2}.
\end{equation}
The corresponding change of the $x$-distance between the beams
$\Delta_{12,x}^1-\Delta_{12,x}^2$ can be deduced from the luminosity change
$L_{LSC}^1-L_{LSC}^2$ and van der Meer scan data.  For example, if one of the
$x$-scans was performed symmetrically, the beam separation change required to
modify the luminosity by a given amount can be calculated from the derivative
$d\Delta_{12,x}/dL = d\alpha^{sym}/dL^{sym}\cdot(a_1+a_2)/2$. This leads to
the equation
\begin{align}
\label{eq:40}
  \Delta_{12,x}^1-\Delta_{12,x}^2&=(\alpha_{LSC}^1-\alpha_{LSC}^2)(a_1-a_2) \nonumber \\
  &= \frac{d\alpha^{sym}}{dL^{sym}}\cdot\frac{a_1+a_2}{2}
(L_{LSC}^1-L_{LSC}^2),
\end{align}
which allows to obtain $(a_1-a_2) / (a_1+a_2)$ from the measurable values
$d\alpha^{sym}/dL^{sym}$, $L_{LSC}^1-L_{LSC}^2$ and the set difference
$\alpha_{LSC}^1-\alpha_{LSC}^2$.  Together with $(a_1+a_2)/2$ from
\cref{eq:102} this allows to calibrate the scales $a_{1,2}$ individually.

To improve the sensitivity of the method and to increase
$L_{LSC}^1-L_{LSC}^2$, LSC can be performed at a point close to the maximum of
the derivative $dL/d\Delta_{12}$, where the second derivative is zero. For
the Gaussian bunches with the widths $\sigma_{1,2}$, the luminosity dependence
on the beam separation is also Gaussian with the sigma
$\Sigma=\sqrt{\sigma_1^2+\sigma_1^2}$, and the optimal LSC beam separation is
$\Delta_{12}=\Sigma$.

After the calibration, the length scale systematics is typically well below
1\%.

Note that LSC is not needed for the ``static'' imaging methods, namely, for
the beam-gas and also for the beam--beam imaging if in the latter the
reconstructed bunch is {\it stationary} during the scan in the laboratory
frame. In both cases, the reconstruction is performed in the vertex detector
coordinates, which always define an accurate scale.

\vspace{5mm}

The remaining most important sources of systematics include $x$-$y$
non-factorizability of the bunch densities, the beam orbit drifts and the
so-called beam--beam effects due to the electromagnetic interaction between
two colliding bunches.

As it was discussed in Sec.~\ref{sec:xy}, the $x$-$y$ non-factorizability can
be circumvented by performing {\it two-dimensional scans} over the central
region giving the main contribution to the integral in \cref{eq:18}. Three
two-dimensional scans, already performed at LHCb at the end of Run 2, allowed
to reduce this uncertainty approximately by an order of magnitude~\cite{twoD}.

The beams can drift at LHC by a few microns leading to the cross-section
uncertainties at the level of 1\%. Further improvements require accurate
monitoring of the beam positions. The LHC Beam Position Monitors (BPMs) could
not provide the necessary accuracy in Run 1 because of the temperature drifts
in the readout electronics. In Run 2 they were upgraded and all interaction
points were equipped with the so-called DOROS BPMs. The accuracy was
significantly improved and reached a sub-micron level. This was proved by
calibrating and comparing with the beam positions reconstructed with the
beam-gas imaging at LHCb. The latter is relatively slow and requires one or a
few minutes to reach the required accuracy even with the gas injection. This
is sufficient for the calibration, however, and after that, the DOROS BPMs can
accurately measure even fast beam drifts with 0.1 second time resolution.
Sufficient accuracy can, possibly, be achieved also at other experiments
without the beam--gas imaging using the correlations between the DOROS
measurements and the positions of the luminous centers. Using well-calibrated
DOROS BPM data, one can significantly reduce the scan-to-scan
non-reproducibility and potentially achieve the overall calibration accuracy
below 1\%.

The last beam--beam systematics is caused by the electromagnetic interaction
between the bunches. The electromagnetic force kicks the beam particles and
modifies their accelerator trajectories and the bunch densities $\rho_{1,2}$.
If the perturbations were constant during the scan it would not produce any
bias since van der Meer formula \cref{eq:vdm} is valid for any densities.  The
kick strength, however, depends on the transverse profile of the opposite
bunch and the distance to it. The perturbation of the densities
$\rho_{1,2}$, therefore, depends on $\Delta x$, $\Delta y$. For example, it
vanishes at large beam separations.  Such $\rho_{1,2}$ dependence on
$\Delta x$, $\Delta y$ breaks the derivation of \cref{eq:vdm} from
\cref{eq:prevdm} and introduces biases in the cross-section formulas that are
difficult to estimate.  The beam--beam correction is the main subject of this
paper and will be discussed in detail in the following sections.

The lead ion bunches in LHC van der Meer scans carry much smaller charge than
the proton ones.  Therefore, the beam--beam systematics is more significant
for the proton--proton scans.

The beam--beam interaction also affects the imaging methods. Like the
classical van der Meer approach, the beam--beam image is biased when
$\rho_{1,2}$ densities vary during the scan. The beam--gas imaging is also
biased when the beam--beam interaction changes, eg. during van der Meer scan at
the same or any other LHC experiment. In the latter case, the associated
beam--beam distortions propagate through the accelerator everywhere in the
ring. If the beams are stationary, however, the densities $\rho_{1,2}$ are
constant and the appropriate $\rho_{1,2}$ fit model can describe the bunches
accurately. The required model might be complicated, however, and dependent on
the constant beam--beam interactions at all LHC experiments.

In the first LHC publications, the experiments either not considered the
beam--beam systematics or assigned $\sim1\%$ error to their luminosity
measurements in the proton--proton scans without making any correction. In
2012 the method~\cite{madx12} was proposed for correcting the classical van
der Meer technique. It was subsequently used by all LHC experiments, and the
systematic error was reduced to
$0.3-0.7\%$~\cite{CMS-PAS-LUM-13-001,Aad:2013ucp,
  lhcb14,Aaboud:2016hhf,cms15,cms16,cms17,atlas19}. However, the beam--beam
force in this method was oversimplified and approximated by a linear function
of the transverse coordinates. More specifically, the electromagnetic field
was described by the dipole and quadrupole magnets responsible for the offset
and the slope of this linear function, respectively.

This was discovered by the author of this paper in January 2019 using a new,
independently developed simulation. It will be presented in this
paper. Instead of the linear approximation, the simulation uses the accurate
formula of the beam--beam force. Unfortunately, the old and new beam--beam
corrections differ by $\sim1\%$ as shown in Fig.~\ref{fig11}. The difference
is dependent on van der Meer scan beam parameters. This requires the
corresponding rescaling of all LHC cross-sections after 2012 that were based
on the luminosity calibrated with the old oversimplified beam--beam model.

The new simulation is primarily oriented at the classical van der Meer
method. It is optimized for calculating the {\it luminosity} but not the {\it
  bunch shapes} required in the imaging methods. Some limited tools for
predicting the shapes are implemented in the simulation, however, and can be
extended.

The beam--beam force depends on both transverse coordinates and, therefore,
introduces $x$-$y$ coupling and non-factorization. The new simulation allows
to correct the luminosity measurements at each point of van der Meer scan such
that the beam--beam perturbation is effectively removed together with its
$x$-$y$ coupling. The cross-section can then be calculated from the corrected
$\mu$ values using unmodified \cref{eq:vdm} or~\cref{eq:fact}.

The new simulation is sufficiently general. The bunch profiles can be
approximated by an arbitrary weighted sum of the Gaussians with the common
center. The $x$- and $y$-widths can be different. In addition, the luminosity
correction can be calculated in the presence of the beam--beam kicks at an
arbitrary number of interaction points. The bunch shapes of all colliding
bunches are specified individually.



\section{Momentum kick induced by the beam--beam interaction} 
\label{sec:beambeam}

For LHC physics one usually considers the collision of two protons (or ions)
ignoring other particles in the bunches. Contrary to that, the beam--beam
electromagnetic interactions have a long-range and act simultaneously between
many particles. At large distances, one may neglect quantum effects and use
classical electrodynamics. Any associated electromagnetic radiation of protons
or ions at LHC will be ignored.

The formula of the momentum kick induced by the beam--beam interaction is well
known in the accelerator community. However, it might be not so easy to find
in the literature its rigorous derivation with the discussion of all
simplifying assumptions affecting the accuracy. To make the material of this
paper self-contained, we present below the derivation of this formula from the
first principles.

\subsection{Electromagnetic interaction of two particles}
As it will be shown in a moment, at LHC the beam--beam force changes the
transverse particle momentum in the laboratory frame by a few MeV,
ie. negligibly compared to the total momentum. Therefore, one can assume that
all particles creating the electromagnetic field move without deflections with
constant velocity. Since the equality and constancy of velocities are preserved
by boosts, this approximation can be used in any other frame.

\begin{figure}[htbp]
\begin{center}
  \includegraphics[width=0.4\textwidth,angle=0]{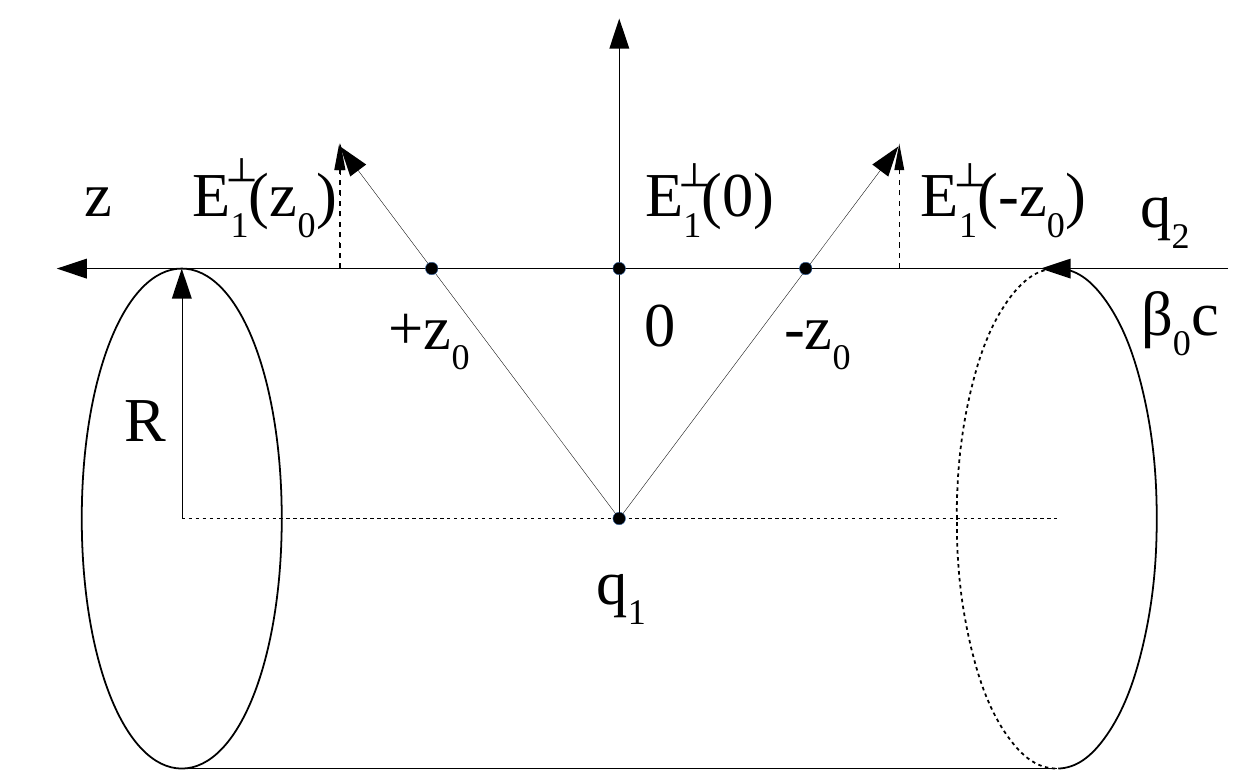}
  \caption{The electrical field from the charge $q_1$ at rest
    acting on the charge $q_2$ from another bunch.}
  \label{fig2}
\end{center}
\end{figure}

The electromagnetic field of a particle with the charge $q_1$ is simplest in
its rest frame where it reduces to the electrical Coulomb component
$\vec{E}_1$ shown in Fig~\ref{fig2}.  The momentum kick exerted on a
particle $q_2$ from another bunch can be calculated in this frame as an
integral of the infinitesimal momentum changes along the trajectory.
According to our assumption, the velocity of $q_1$ is constant, so its
position is fixed.  Since $q_2$ in the rest frame of $q_1$ has even larger
momentum than in the laboratory, while the transverse kick is the same, one can
safely assume that its speed $dz / dt = \beta_0 c$ is also constant and $q_2$
moves along the straight line denoted as the $z$-axis in Fig.~\ref{fig2}.

If the velocities of the colliding particles in the laboratory system are
$\pm\vec{\beta} c$, $\beta_0$ can be expressed as
\begin{equation}
\label{eq:105}
\beta_0 = 2\beta/(1+\beta^2),
\end{equation}
which is the double-angle (or double-rapidity) formula for the hyperbolic
tangent, the analog of $\tan(2\phi)=2\tan\phi/(1-\tan^2\phi)$. Of course, at
LHC one can safely assume $\beta \approx \beta_0 \approx 1$.

The momentum kick received by $q_2$ is given by the line integral
\begin{equation}
  \label{eq:kickint}
  \Delta \vec{p}_2 = \int q_2 \vec{E}_1^\perp (x,y,z) \, dt = \int q_2 \vec{E}_1^\perp (x,y,z) \,
  \frac{dz}{\beta_0 c}.
\end{equation}
Because of the reflection invariance $z\to-z$, the $z$-projection of the kick
should vanish, so only the perpendicular component of the electric field is
written in \cref{eq:kickint}. Its integral
$\int \vec{E}_1^\perp (x,y,z) \, dz$ depends only on the transverse
coordinates. It will be denoted in the following simply by
$\vec{E}_1(x,\, y)$.  According to Gauss's flux theorem applied to the
cylinder with the radius $R$ shown in Fig.~\ref{fig2}, it is
equal to
\begin{equation}
  \label{eq:gausslaw}
  E_1(x,\, y) = \int E_1^\perp (x,y,z) \, dz = \frac{q_1}{2\pi R \epsilon_0},
\end{equation}
where $\epsilon_0$ is the electric constant. Here, we are using the system of
units where the first Maxwell's equation is written as
$\vec{\mathbf{\nabla}} \cdot \vec{E} = \rho / \epsilon_0$, like the
International System of Units (SI).

Substituting \cref{eq:gausslaw} to \cref{eq:kickint} gives
\begin{equation}
  \label{eq:kicksummary}
  \Delta p_2 = \frac{q_2 E_1(x,\, y)}{\beta_0 c} = \frac{q_1 q_2}{2\pi R \epsilon_0 \beta_0 c} =
  \frac{2\alpha \hslash Z_1 Z_2}{R\beta_0},
\end{equation}
where \(\alpha = e^2 / (4\pi c \hslash \epsilon_0)\) is the fine-structure
constant, $e$ is the proton charge, $Z_{1,2} = q_{1,2} / e$ are
either 1 or 82 for proton or lead ion LHC beams, respectively, and $\hslash$
is the reduced Plank constant.

Since the momentum kick $\Delta p_2$ and $R$ are perpendicular to the
$z$-axis, they are conserved by the boosts along this line. Therefore, the
corresponding terms in \cref{eq:kicksummary} remain invariant and equal in all
frames where $q_{1,2}$ velocities are parallel. The perpendicular electric
field $E_1(x,\, y)$, however, depends on the boosts and acquires a magnetic
counterpart. In \cref{eq:kicksummary} it should be used only in the $q_1$ rest
frame.

If $q_1$ and $q_2$ collide in the laboratory system with the small crossing
angles $\alpha_{1,2}$ as in Fig.~\ref{fig1}, \cref{eq:kicksummary} receives
corrections of the order $\alpha_{1,2}^2$. For example, they can be calculated
by boosting the kick \cref{eq:kicksummary} from the primed to the laboratory
system with the velocity $\vec{v_{\perp}}$ and the subsequent projection to
the planes transverse to the beams. At LHC these corrections are negligible.

Equation \cref{eq:kicksummary} is sufficient for a rough estimation of the
momentum kick induced by the whole bunch. If its charge $\sim\!10^{11}e$ is
condensed into $q_1$, the kick at, for example, one bunch sigma
$\sim 100\,\mu m$ is equal to $\Delta p=3$~MeV$/c$. This is, indeed, negligible
compared to LHC energies. The exact formulas of the Gaussian bunch fields are
presented later in \cref{eq:2} and~\cref{eq:BassettiErskine} and lead to the
same conclusion.

Equation \cref{eq:kicksummary} was derived for the stationary charge
$q_1$. Due to the momentum conservation, however, it receives the opposite
kick $\Delta \vec{p}_1=-\Delta \vec{p}_2$ and starts moving. Our formulas do
not depend on the mass of $q_1$ and, therefore, should be valid even when the
mass is much less than $\Delta p_1\sim3$~MeV$/c$. But then after the kick
$q_1$ becomes ultrarelativistic, and its field can not be described by the
simple electrostatics.

To solve this seeming contradiction one should recall that the field from the
ultrarelativistic charge $q_2$ in Fig.~\ref{fig2} is concentrated only in a
thin ``pancake'' perpendicular to $z$ and travelling together with the particle.
So, $q_1$ receives the kick $\Delta p_1$ when $q_2$ passes $z=0$. Only after
that and almost instantaneously $q_1$ becomes ultrarelativistic. Let's denote
this moment by $t(z=0)$. It then takes some time to propagate this information
back to $q_2$, which is escaping almost at the speed of light.  Namely, $q_2$
``sees'' the initial stationary field from $q_1$ until it passes the forward
light cone emitted from $q_1$ at $t(z=0)$. This moment $t(z)$ can be found
from
\[(t(z)-t(z=0))^2 =(z/c\beta_0)^2=(R^2+z^2)/c^2,\]
so $q_2$ travels the distance
$z=R\beta_0\gamma_0\gg 1$ where $\gamma_0=(1-\beta_0)^{-1/2}$. Therefore, most
of $\Delta p_2$ kick is created by the stationary $q_1$.

From this consideration we see again that the assumption of constant $q_{1,2}$
velocities is valid only in the {\it ultrarelativistic} limit.  Therefore, the
dependence $1/\beta_0$ in \cref{eq:kicksummary} is not justified and should be
dropped. The formula should be written as
\begin{equation}
\label{eq:7}
  \Delta p = \frac{q_2 E_1(x,\, y)}{c} = \frac{2\alpha \hslash Z_1 Z_2}{R}
\end{equation}
under the explicit condition $\beta_0\approx1$. The corresponding angular kick
for the particles with the momentum $p$ is
\begin{equation}
\label{eq:8}
\Delta\phi = \frac{q_2 E_1(x,\, y)}{pc} = \frac{2\alpha \hslash Z_1 Z_2}{Rp}.
\end{equation}

Note that in the present literature the angular kick and the related
parameters traditionally and most often are expressed via the classical
particle radius \(r_c = q^2/(4\pi \epsilon_0 mc^2)\) determined by the
particle mass $m$. For example, one can refer to the so-called {\it beam--beam
  parameter}.  In the recent Particle Data Group
review~\cite{10.1093/ptep/ptaa104} it is defined in Eq.~(31.13) as
\[
  \xi_{y2} = \frac{m_er_{e}q_1q_2N_1\beta_{y2}}{2\pi m_2\gamma_2\sigma_{y1}(\sigma_{x1}+\sigma_{y1})}
\]
for the $y$-direction, where $\beta_{y2}$ is the beta-function discussed
later, $m_2$, $\gamma_2$ are the mass and $\gamma$-factor of $q_2$ and $m_e$,
$r_e$ are the electron mass and classical radius. Writing mass in the formulas
is misleading, since, as it was explained above, the kick {\it does not}
depend on $m$. In the ultrarelativistic case $\beta_0 \approx 1$ the momentum
change $\Delta p$ depends {\it only} on the electric charges $q_{1,2}$ and
$R$, since it is determined by Gauss's or Coulomb's law.  If the mass $m$ is
introduced in the beam--beam equations, it should necessarily cancel as in the
combination $m_er_e$ above.  For example, the kick is the same for protons and
electrons if their momenta are the same, despite the difference in their
classical radii.  To stress this invariance, the fine-structure constant
\(\alpha\) should be used instead of the classical radius because only the
electromagnetic interaction is relevant here. It is better to drop completely
the mass $m$ from the formulas.

\subsection{Simplifying assumptions in the particle interaction with the
  opposite bunch}
\label{rigidbunch}
Let's demonstrate that for calculating the kick from the whole bunch one can
assume that all bunch particles move in the same direction with the same
speed. As it will be discussed in Sec.~\ref{accdynam}, the angular spread in
the laboratory frame is of the order
$\delta\alpha=\sigma_T/\beta\lesssim O(10^{-5})$ where $\sigma_T$ is the
transverse bunch size ($40-100\,\mu m$) and $\beta$ is the beta-function in
the range $1- 20$~m during van der Meer scans in the interaction points at
LHC. Note also that the kick is perpendicular to the velocity difference
$\Delta\vec{v}=\vec{v}_1-\vec{v}_2$ from Fig.~\ref{fig1}, so the kick angular
variation is of the same negligible order $\delta\alpha$.

There is one effect where this spread is enhanced.  An angular deviation of
one particle changes its crossing angle with respect to the opposite
bunch. This affects the transverse bunch width $\sigma_T'$ {\it visible from the
particle} according to \cref{eq:27}. Therefore, the angular
variation $\delta\alpha$ leads to the effective smearing of $\sigma_T'$:
\begin{equation}
\label{eq:30}
\frac{\delta\sigma_T'}{\sigma_T'} \approx
\left(\frac{\sigma_L}{\sigma_T}\right)^2\alpha\ \delta\alpha =
\left(\frac{\sigma_L}{\sigma_T}\right)
\left(\frac{\sigma_L}{\beta}\right)\alpha.
\end{equation}
In spite of the large enhancement factor $\sigma_L/\sigma_T\lesssim1000$, the
values of $\sigma_L/\beta\lesssim O(10^{-2})$ and $\alpha\lesssim O(10^{-4})$
are so small that in van der Meer scans the variations of the transverse width
$\delta\sigma_T'/\sigma_T' \lesssim 0.01$ can be neglected in the beam--beam kick
calculations.

The longitudinal momentum spread $\delta p/p$ of the beam is completely
negligible for our purposes, since the beam--beam kick is determined by the
velocities that are close to the speed of light at LHC and almost insensitive
to the momentum change. Namely, if $v$ is the velocity corresponding to the
rapidity $\phi$, $v=\tanh\phi$, its change is
$\delta v=\delta\phi/\cosh^2\phi=\delta(\sinh\phi)/\cosh^3\phi=(\delta
p/p)\cdot\beta/\gamma^2\propto 1/\gamma^2$.  The associated angular variation
of $\Delta\vec{v}=\vec{v}_1-\vec{v}_2$ due to the crossing angle is
additionally suppressed by the smallness of $\alpha< 10^{-3}$.

Finally, the angular variation due to the beam--beam kick itself is also
small, $\Delta p/p\sim 10^{-6}$. Since the typical longitudinal bunch length
$\sigma_L$ is 5--10~cm, the kick has no time to develop to a sizable
displacement {\it during} the interaction.  The particles should travel freely
much longer distances of the order $\sigma_T \cdot p/\Delta p\sim100$~m before
their transverse displacements reach $\sigma_T$. However, the accelerator
elements controlling the transverse movements correct the trajectories and
bring the particles back. In Sec.~\ref{xchecks} equation \cref{eq:121} it will
be shown that in the end the beam orbit is shifted by less than $1\%$ of the
bunch width. The {\it angular} distribution shifts by
$\Delta p/2p\lesssim 10^{-6}$. The beam--beam luminosity bias typically does
not exceed 1\%. Therefore, to achieve the required overall calibration
accuracy of 0.1\% and to estimate the bias with the relative uncertainty
$<0.1\%/ 1\% = 10\%$, it is sufficient to calculate the momentum kicks using
the electromagnetic fields of the {\it unperturbed densities} $\rho_{1,2}$.

If one can assume that the particles in the bunches move with constant and
opposite velocities, this greatly simplifies our four-dimensional
electromagnetic problem and reduces it to the two-dimensional
electrostatics. Indeed, in \cref{eq:gausslaw} one can easily recognize the
Coulomb's law in two dimensions.  The circle circumference $2\pi R$ in the
denominator substitutes the sphere area $4\pi R^2$ in the three-dimensional
Coulomb's law in accordance with the Gauss's electric flux theorem. Therefore, 
\begin{equation}
\label{eq:12}
\vec{F}_{12} = \Delta \vec{p}_1c = q_1\vec{E}_2 = -q_2\vec{E}_1 = -\Delta \vec{p}_2c
\end{equation}
from \cref{eq:7} is just the Coulomb's two-dimensional force between $q_1$ and
$q_2$. The calculation of $\vec{\Delta} p_{1,2}$ kick, ie.  the problem of the
electromagnetic interaction of the ultrarelativistic laboratory bunches,
reduces to the calculation of the two-dimensional electrostatic forces between
the transversely projected {\it static} charges in the frame with the parallel
beams.

As it was already discussed, the {\it longitudinal} bunch distributions do not
matter in this frame. Indeed, the accumulated kick remains invariant if
particles in the opposite bunch are arbitrarily displaced longitudinally as
long as they follow the same lines and traverse the whole interaction region.

\subsection{Electrostatic field from two-dimensional Gaussian distribution}
In this subsection we present the formulas of the electrostatic field from the
two-dimensional Gaussian density
\begin{equation}
\label{eq:1}
\rho = \frac{Q}{2\pi\sigma_x\sigma_y}\exp\left(-\frac{x^2}{2\sigma_x^2}-\frac{y^2}{2\sigma_y^2}\right).
\end{equation}
For the round bunch with $\sigma_x=\sigma_y=\sigma$, the azimuthally symmetric
field can be determined from the charge $Q\left(1-e^{-R^2/2\sigma^2}\right)$
inside the disk $x^2+y^2<R^2$ and the Gauss's flux theorem:
\begin{equation}
\label{eq:2}
E = \frac{Q}{2\pi\epsilon_0 R}\left(1-e^{-R^2/2\sigma^2}\right).
\end{equation}
Therefore, the beam--beam angular kick of the particle with the charge $Z_1e$
induced by the round Gaussian bunch with $N_2$ particles with the charges
$Z_2e$ is
\begin{equation}
\label{eq:13}
\Delta\phi=\frac{Z_1eE}{pc}=\frac{2\alpha\hslash Z_1Z_2N_2}{pR}\left(1-e^{-R^2/2\sigma^2}\right).
\end{equation}

The field from an elliptical bunch with $\sigma_x\ne\sigma_y$ is more
complicated. It was derived by Bassetti and Erskine in~\cite{Bassetti:122227}:
\begin{equation}
\label{eq:BassettiErskine}
 E_x-iE_y  =
 \frac{Q \cdot e^{-z_2^2}}{\pi\epsilon_0\sqrt{2\left(\sigma_x^2-\sigma_y^2\right)}}
   \int _{z_1}^{z_2}
e^{\zeta^2}\,d\zeta,
\end{equation}
where the path-independent integral is taken in the complex plane between the
points
\begin{equation}
\label{eq:16}
z_1 = \frac{x\frac{\sigma_y}{\sigma_x} + iy\frac{\sigma_x}{\sigma_y}}{\sqrt{2\left(\sigma_x^2-\sigma_y^2\right)}},\quad
z_2 = \frac{x+iy}{\sqrt{2\left(\sigma_x^2-\sigma_y^2\right)}}.
\end{equation}
It can be expressed via the complex error function $\mathrm{erf}(z) = 2\int_0^ze^{-\zeta^2}\,d\zeta/\sqrt{\pi}$
or its scaled version named Faddeeva function
\begin{equation}
\label{eq:3}
w(z)=e^{-z^2}\left(1+\frac{2i}{\sqrt{\pi}}\int_0^ze^{\zeta^2}d\zeta\right)
\end{equation}
as
\begin{equation}
\label{eq:31}
 E_x-iE_y  =
 -iQ\frac{w(z_2) -
   w(z_1)\exp\left(-\frac{x^2}{2\sigma_x^2}-\frac{y^2}{2\sigma_y^2}\right)}{2\epsilon_0\sqrt{2\pi(\sigma_x^2-\sigma_y^2)}}.
\end{equation}
Note that in~\cite{Bassetti:122227} the sign in front of $y^2/2\sigma_y^2$ was
misprinted as plus.  A simplified proof of Bassetti--Erskine formula found by
the author will be published in a separate paper.

The Faddeeva function $w(z)$ grows exponentially when the imaginary part
$Im(z)$ of its argument tends to $-\infty$. In this case, calculating the
difference between two large numbers in the enumerator of \cref{eq:31}
becomes numerically unstable. In practice, to ensure the positiveness of
$Im(z_{1,2})$, the calculation can be performed in the following way.  In the
case $\sigma_x<\sigma_y$ the formulas might be applied with the {\it swapped}
$x$- and $y$-directions.  The obtained components $E_x$ and $E_y$ should be
swapped back. This ensures that the square roots in \cref{eq:16} are always
taken with $\sigma_x>\sigma_y$ and, therefore, are real.  Then $Im(z_{1,2})$
becomes negative only if $y<0$. Since the field is centrally symmetric,
$\vec{E}(x,y)=-\vec{E}(-x,-y)$, this case can be circumvented by calculating
the field at the opposite point $(-x,-y)$ and by inverting the signs of the
obtained components $E_x$, $E_y$.

\subsection{Average kick of bunch particles}
\label{avrkick}
Up to now, we have discussed the kicks of individual particles. In this
subsection we present a simple formula for the kicks averaged over the
bunches.  For the Gaussian shapes, it was derived in Appendix~A
in~\cite{Hirata:1987cn}. Here we give an alternative proof based on simple
arguments and extend the formulas to arbitrary $\rho_{1,2}$.

Let's denote the momentum kick of the $i$-th particle in the first bunch
exerted by the $j$-th particle in the second by $\Delta \vec{p}_{ij}$. It can
be calculated as $\Delta \vec{p}_{ij}=\vec{F}_{ij}/c$ where
\begin{equation}
\label{eq:36}
\vec{F}_{ij} = \frac{q_iq_j}{2\pi\epsilon_0}\frac{\Delta\vec{r}_{ij}}{|\Delta\vec{r}_{ij}|^2}
\end{equation}
is the two-dimensional Coulomb's force between the charges $q_i$, $q_j$
separated by $\Delta\vec{r}_{ij} = \vec{r}_i-\vec{r}_j$ in the transverse
plane.  The full force on the first bunch is the sum $\sum_{i,j}\vec{F}_{ij}$
that can be approximated by the integral
\begin{equation}
\label{eq:33}
\vec{F}_1=\sum_{i,j}\vec{F}_{ij} = N_1N_2\int \vec{F}(\vec{r}_1-\vec{r}_2)\rho_1(\vec{r}_1)\rho_2(\vec{r}_2)\,d\vec{r}_1\,d\vec{r}_2,
\end{equation}
where $N_{1,2}$ are the number of particles in the bunches.  Since $\vec{F}$
depends only on the difference $\Delta\vec{r} = \vec{r}_1-\vec{r}_2$, it is
convenient to use $\Delta\vec{r}$ as the integration variable 
\begin{equation}
\label{eq:34}
\vec{F}_1=
 N_1N_2\int \vec{F}(\Delta\vec{r})\rho(\Delta\vec{r})\,d\Delta\vec{r},
\end{equation}
where $\rho$ also depends only on $\Delta\vec{r}$:
\begin{equation}
\label{eq:35}
\rho(\Delta\vec{r}) = \int\rho_1(\Delta\vec{r}+\vec{r}_2)\rho_2(\vec{r}_2)\,d\vec{r}_2.
\end{equation}
This is the cross-correlation $\rho_2\star\rho_1$ or, equivalently, the
convolution $\rho_1*\tilde\rho_2$ where
$\tilde\rho_2(\vec{r}) = \rho_2(-\vec{r})$ ie. $\rho_2(\vec{r})$ with an
opposite argument.

Equations \cref{eq:33} and \cref{eq:34} show that $\Delta\vec{r}$ spread in
the integral can be equivalently represented either by the two bunch densities
$\rho_{1,2}$ or by only one $\rho$. In the latter case the second bunch
effectively collapses to the point-like charge at the origin. Indeed,
\cref{eq:34} follows from \cref{eq:33} if $\rho_1$ and $\rho_2$ are
substituted by the artificial bunch density $\rho=\rho_1*\tilde\rho_2$ and by
the delta-function at zero, respectively.

\begin{figure}[htbp]
\begin{center}
  \includegraphics[width=0.155\textwidth,angle=0]{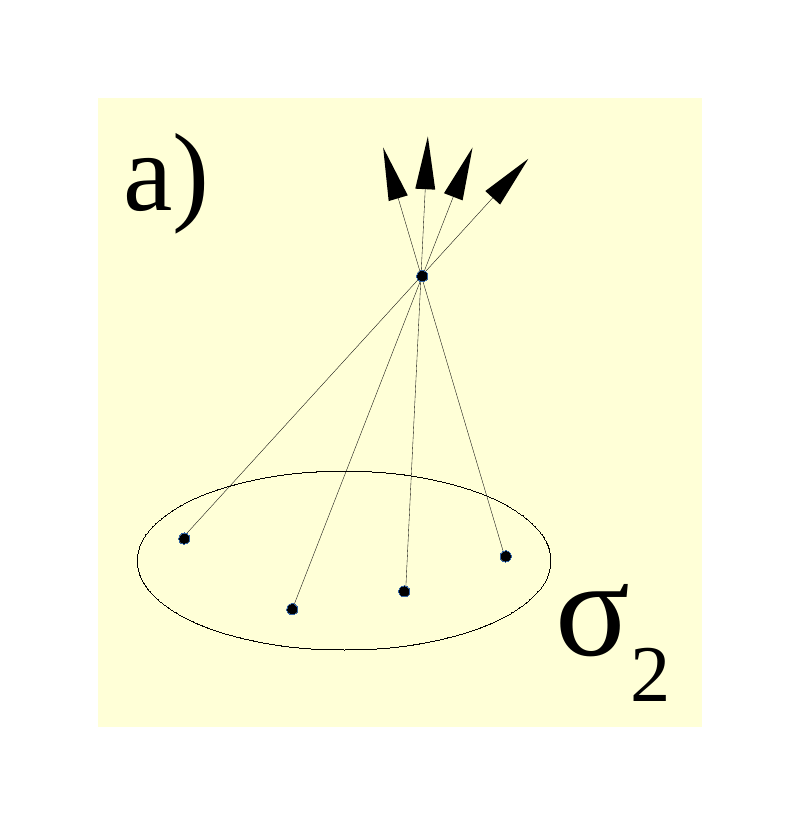}
  \includegraphics[width=0.155\textwidth,angle=0]{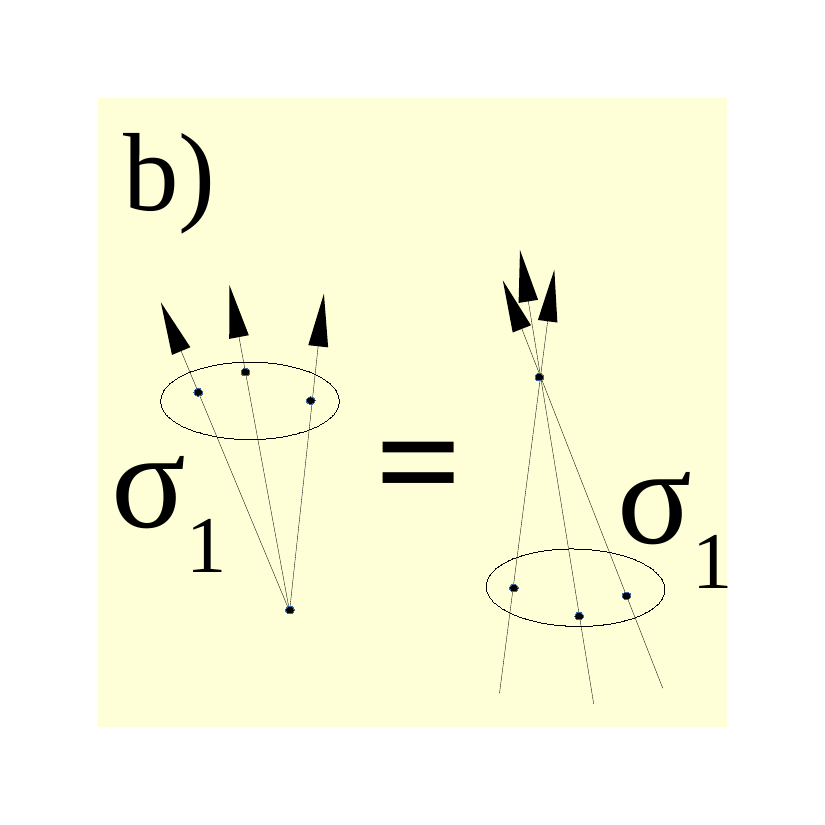}
  \includegraphics[width=0.155\textwidth,angle=0]{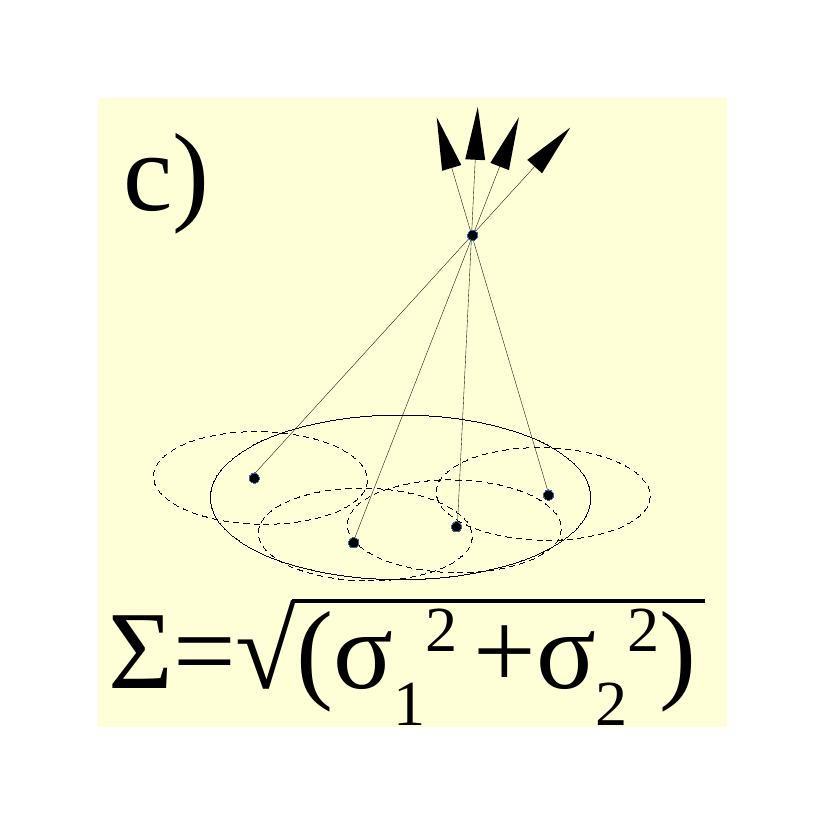}
  \vspace{-2mm}
  \caption{ The force between the bunches does not change when one bunch
    density $\rho_i(\vec{r})$ ($i=1,2$) is collapsed to the point charge at
    the origin while the other is convolved with $\rho_i(-\vec{r})$. For the
    Gaussian $\rho_{1,2}$ densities the convolution has sigma
    $\Sigma=\sqrt{\sigma_1^2+\sigma_2^2}$.}
  \label{fig3}
\end{center}
\end{figure}

This is illustrated schematically in Fig.~\ref{fig3}. The left picture shows
the overall electrostatic force $\sum_j\vec{F}_{ij}$ exerted by the second
bunch on the $i$-th particle. The {\it origins} of $\vec{F}_{ij}$ vectors are
varied according to the density $\rho_2(x,y)$.  To obtain the full force, one
needs to sum over $i$, ie. to vary the {\it ends} of $\vec{F}_{ij}$ vectors.
Fig.~\ref{fig3}b shows such variations for the $j$-th particle of the second
bunch.  Since parallel translations do not change vectors $\vec{F}_{ij}$, the
variations of their {\it ends} can be obtained by the {\it opposite}
variations of the origins, as shown in the picture. The full sum
$\sum_{i,j}\vec{F}_{ij}$ in Fig.~\ref{fig3}c, therefore, can be calculated by
smearing the {\it origins} with both probability densities $\rho_1(-\vec{r})$
and $\rho_2(\vec{r})$. Equivalently, it can be calculated by varying the {\it
  ends} of $\Delta\vec{r}$ by $\rho=\rho_1(\vec{r}) * \rho_2(-\vec{r})$ in
agreement with \cref{eq:35}. For the Gaussian bunches the convolution
$\rho_1*\tilde\rho_2$ is again Gaussian with the sigmas
$\Sigma_{x,y}=\sqrt{\sigma_{1x,y}^2+\sigma_{2x,y}^2}$.

Since the momentum is conserved, the full momentum kicks of two bunches are
opposite, $\vec{F}_1=-\vec{F}_2$. The average kicks of the particles are equal
to $F_1/N_1$, $F_2/N_2$ and are different if $N_1\ne N_2$.

\section{Beam--beam numeric simulation B*B}

Unfortunately, there is no known analytic method that can predict the
luminosity change caused by the beam--beam effect. Below we describe a new
numerical simulation developed for this purpose named ``B*B'' or ``BxB''
(pronounced ``B-star-B'')~\cite{BxB1},~\cite{BxB2}. Before going into details,
let's briefly remind the transverse dynamics of the particles in an idealized
accelerator, the so-called ``betatron motion''.

\subsection{Recurrence relation}
\label{accdynam}
Every particle in the beam oscillates around a stable orbit with a constant
amplitude. Ideally, the oscillations in $x$ and $y$ are independent. They are
described by the Hill's equation
\begin{equation}
\label{eq:51}
u'' + K_u(s) u(s) = 0,
\end{equation}
where $u$ is the transverse coordinate ($x$ or $y$), $s$ is the circular
coordinate along the ring, $u'' = \partial^2 u/\partial s^2$ and $K_u(s)$ is a
function defined by the quadrupole accelerator elements whose field is
proportional to $u$. The solutions of \cref{eq:51} are
\begin{equation}
\label{eq:41}
u = \sqrt{\epsilon_u\beta_u(s)} \cos(\phi_u(s)-\phi_{0,u}),
\end{equation}
where $\epsilon_u$ is a constant defining the oscillation amplitude and called
``emittance'' in the accelerator language, $\beta_u(s)$ is the so-called
``beta-function'' determined by the equation
\begin{equation}
\label{eq:52}
\frac{1}{2}\beta_u\beta_u'' - \frac{1}{4}\beta_u'^2 + \beta_u^2 K_u = 1,
\end{equation}
while
\begin{equation}
\label{eq:53}
\phi_u(s) = \int_0^s \frac{d\zeta}{\beta_u(\zeta)}.
\end{equation}
is the ``phase advance'' whose value at $s=0$ is denoted by $\phi_{0,u}$. From
\cref{eq:41} and~\cref{eq:53} one can calculate $u' = \partial u/\partial s$,
ie. the tangent of the angle between the particle and the orbit:
\begin{equation}
\label{eq:68}
u' = \sqrt{\frac{\epsilon_u}{\beta_u}}\left(\frac{\beta_u'}{2}\cos(\phi_u-\phi_{0,u})- \sin(\phi_u-\phi_{0,u})\right).
\end{equation}
At the interaction point the beams are maximally focused to reach the maximal luminosity. As it follows
from \cref{eq:41}, $\beta_u$ is then minimal and $\beta_u'=0$, so that \cref{eq:68} simplifies to
\begin{equation}
\label{eq:77}
u' = -\sqrt{\frac{\epsilon_u}{\beta_u}} \sin(\phi_u-\phi_{0,u}).
\end{equation}
After every turn in the accelerator the particle phase advance increases by the constant 
\begin{equation}
\label{eq:80}
Q_u = \frac{1}{2\pi}\int_0^L \frac{d\zeta}{\beta_u(\zeta)}
\end{equation}
called the ``tune''. Here, the integral is taken over the whole
accelerator length $L$.  As it follows from~\cref{eq:41} and~\cref{eq:77}, at
the interaction point it is convenient to merge the phase-space coordinates
$(u,u')$ to one complex variable
\begin{equation}
\label{eq:78}
z_u = u - iu'\beta_u.
\end{equation}
Its evolution is described by the simple rotation in the complex plane
\begin{equation}
\label{eq:79}
z_{n+1,u} = \sqrt{\epsilon_u\beta_u} e^{i(\phi_{n+1,u}-\phi_{0,u})} = z_{n,u}e^{2\pi iQ_u},
\end{equation}
where $z_{n+1,u}$ and
$z_{n,u}=\sqrt{\epsilon_u\beta_u} e^{i(\phi_{n,u}-\phi_{0,u})}$ are the
complex coordinates at the turns $n+1$ and $n$, respectively. Note that the
minus sign in \cref{eq:78} is chosen according to the minus in \cref{eq:77},
so that the rotation is counter-clockwise by definition. Since there are two
transverse axes $x$ and $y$, there are independent rotations in the two
complex planes $z_x=x-ix'\beta_x$, $z_y=y-iy'\beta_y$ and the full phase-space
is four-dimensional.

The bunch transverse shapes are typically approximated by Gaussians. The
distributions in every complex plane then have a form of the two-dimensional
Gaussians with the same standard deviations along $u$ and $-u'\beta_u$. They
are invariant under rotations around the origin and transform to themselves
after every accelerator turn.

As it was discussed, the beam--beam kick changes the angle $u'$ while the
instantaneous change of $u$ is negligible. Equation \cref{eq:79} then modifies
to the recurrence relation
\begin{equation}
\label{eq:81}
z_{n+1,u} = (z_{n,u} - i\beta_u \Delta u')e^{2\pi iQ_u}.
\end{equation}
According to \cref{eq:12}, the angular kicks in the first bunch, for example,
are determined by $(\Delta x',\,\Delta y')=q_1\vec{E}_2(x,y)/pc$. The
electrostatic field is given by \cref{eq:2} or~\cref{eq:BassettiErskine} for
round and elliptical bunches, respectively.  The beam--beam deformations of the
bunch creating the field are neglected, as it was explained in
sect.~\ref{rigidbunch}.

The strategy of the B*B simulation is, therefore, the following. In the
beginning, the particles are distributed in the phase-space according to the
given initial density. Then, they are propagated through the accelerator
turn-by-turn using \cref{eq:81} and the change of the luminosity integral
\begin{equation}
\label{eq:83}
\int (\rho_1+\delta\rho_1)\rho_2dx\,dy
\end{equation}
is calculated. To take into account the beam--beam perturbation of the second
bunch shape, the simulation is repeated with the swapped bunches yielding
$\int \rho_1(\rho_2+\delta\rho_2)\,dy$. The full luminosity change with
respect to the unperturbed value is approximated as
\begin{align}
  \label{eq:82}
  \int \left(\rho_1+\delta\rho_1\right)\left(\rho_2+\delta\rho_2\right)dx\,dy - \int \rho_1\rho_2dx\,dy  \nonumber\\
  \approx
  \int \left(\delta\rho_1\cdot\rho_2 +\rho_1\cdot\delta\rho_2\right) dx\,dy.
\end{align}
The second-order term $\int \delta\rho_1\delta\rho_2dx\,dy$ is neglected.  If
the bunches are identical, two terms in \cref{eq:82} coincide and it is
sufficient to perform one simulation and to double the correction.

The main challenge of the numerical calculation of the beam--beam modified
luminosity \cref{eq:83} is the required accuracy. It should be negligible compared to other
systematic uncertainties, ie. less than $0.1\%$ at LHC.  Reaching this level
with the Monte Carlo integration in the four-dimensional phase space
requires simulating many particles and too much CPU time. Therefore, several
optimizations are implemented in B*B that are described below.

\subsection{Particle weights}
In \cref{eq:82} only the integrals of perturbed and unperturbed densities are
required. Contrary to the unperturbed profile defined in B*B by a continuous
analytic formula, the other density, eg.  $\rho_1+\delta\rho_1$ in
\cref{eq:83}, should be represented by the point-like particles. This can be
achieved by splitting the full phase-space into volumes $V_i$, $i=1,2,\ldots$
. Each of them can be assigned to one ``macro-particle'' with the weight $w_i$
equal to the phase-space density integrated over $V_i$ ,
$w_i=\int_{V_i} (\rho_1+\delta\rho_1)d^4V$, where $d^4V=dx\,dx'\,dy\,dy'$.  In
this way any continuous density can be approximated by a weighted sum of
delta-functions placed at the macro-particle coordinates.  The integral from
\cref{eq:83} can then be expressed as the discrete sum
\begin{align}
\label{eq:84}
  \int (\rho_1+\delta\rho_1)\rho_2dx\,dy \approx
  &\sum_{i=1}^N \rho_2(x_i,y_i)\int_{V_i} (\rho_1+\delta\rho_1)d^4V \nonumber\\
=&\sum_{i=1}^Nw_i\rho_2(x_i,y_i).
\end{align}
Here, the density $\rho_2$ is considered constant in $x$-$y$ projection of
each $V_i$ and substituted by its value $\rho_2(x_i,y_i)$ at the particle
position $(x_i,y_i)$.

Let's assume that $N_1$ real particles in the first bunch are approximated by
$N_1^{B*B}\ll N_1$ macro-particles. Let's define that the association of the
real and macro-particles does not change, so that each $V_i$ always contains
the same $N_1\int_{V_i}(\rho_1+\delta\rho_1)d^4V=N_1w_i$ particles.  With
this definition, the volumes $V_i$ deform due to the beam--beam force but the
weights $w_i$ are conserved as it follows from the conservation of particles.
To simplify notations, the simulated macro-particles in the following
discussion will also be called ``particles'' as the meaning will be clear from the
context.

Since $w_i$ are conserved, in B*B they are defined using the simplest
unperturbed density $\rho_1$ as explained below. To ensure a good sampling of
the four-dimensional space, B*B adopts a two-step approach.  Initially, the
macro-particles are distributed at the circles whose radii $r_{x,y}$ form an
equidistant grid
\begin{equation}
\label{eq:86}
r_x^i=(n_x^i-0.5)\Delta_x, \quad r_y^i=(n_y^i-0.5)\Delta_y,
\end{equation}
where $n_{x,y}^i=1,2,\ldots,n_{max}$. The index $i$ runs across all
$1,2,\ldots,n_{max}^2$ radii pairs $(r_x,r_y)$, where
\begin{equation}
\label{eq:91}
n_{max}^2=N_{part}\sim O(1000)
\end{equation}
is the configurable parameter of the simulation. Each pair receives {\it one}
macro-particle placed {\it randomly} at the circles in the $z_x$, $z_y$
planes. In this way the sampling of {\it the absolute values} $|z_x|, |z_y|$
is realized. The sampling of $z_{x,y}$ {\it phases} is performed by the
accelerator simulation itself. After every turn the particle gets rotated by
$2\pi Q_{x,y}$ angles in the corresponding planes around the origin and
slightly shifted vertically by the beam--beam kick according to
\cref{eq:81}. Since the beam--beam interaction is small at LHC, the particle
trajectories remain approximately circular.  After $N_{turn}^{BB}\sim O(100-1000)$
accelerator turns the particle well samples its circular trajectory and $z_x$,
$z_y$ phases. It was proved that the initial choice of random phases has
negligible impact on the final integral. Equation \cref{eq:81} is applied
approximately $N_{part}\times N_{turn}^{BB}=O(10^6-10^7)$
times and all calculated coordinates contribute to the sampling and the final
integral.  The luminosity sum over all particles \cref{eq:84} is calculated in
B*B after every turn. The average gives the final result.

In principle, it is possible to assign initially not one but several particles
to the $i$-th pair of $z_{x,y}$-circles with the radii $(r_x^i,r_y^i)$. This
would reduce the phase sampling dependency on the tunes and the associated
evolution in the accelerator. The choice of one particle per circle in B*B was
made to simulate more accelerator turns using the same number of
calculations. This allows checking that no new effects appear after a very
large number of turns.

For the normally distributed density in the complex plane $z_u=u-i\beta_uu'$,
\begin{equation}
\label{eq:85}
\rho_1^u(z_u)=\frac{1}{2\pi\sigma_u^2}\exp\left(-\frac{|z_u|^2}{2\sigma_u^2}\right),
\end{equation}
where $u=x,y$, and the equidistant radii $r_u$ from \cref{eq:86}, the
macro-particle weights $w_i$ are given by the integrals over the rings
$n_u^i\Delta_u<r_u<(n_u^i+1)\Delta_u$:
\begin{equation}
\label{eq:87}
w_i^u=\iint_i\rho_1^ud\phi_u\,dr_u\approx\frac{1}{\sigma_u^2}\exp\left(-\frac{(r_u^i)^2}{2\sigma_u^2}\right)r_u^i\Delta_u.
\end{equation}
The final weight of the particle placed at the radii $(r_x^i,r_y^i)$ is the
product
\begin{equation}
\label{eq:89}
w_i=w_i^xw_i^y .
\end{equation}

In van der Meer scans at LHC, the Gaussian bunch approximation is not always
sufficient.  The $x,y$-projections sometimes can be better described by the
sum of two Gaussians. For flexibility, B*B simulation allows defining
$\rho_{1,2}$ as the sum of an {\it arbitrary} number of Gaussians with
configurable weights and widths and independently in $x$ and $y$.
The field $\vec{E}(x,y)$ is calculated from each Gaussian individually using
\cref{eq:2} or~\cref{eq:BassettiErskine}. After weighing, all contributions
are summed.  To speed up the simulation, the field map is precalculated in the
beginning, and then the interpolations are used. This is especially important
in the case $\sigma_x\ne\sigma_y$ when the field should be computed using the
complicated Bassetti-Erskine formula \cref{eq:BassettiErskine}.  Similarly,
both the initial weights $w_i$ and the densities $\rho_2(x_i,y_i)$ at every
turn appearing in \cref{eq:84} are calculated as weighted sums of the
contributions from all Gaussians.

The radii limits $n_{max}\Delta_x$, $n_{max}\Delta_y$ in \cref{eq:86} in the
multi-Gaussian case are chosen in the B*B simulation such that they have the
same efficiency as $r_x<N_{\sigma}\sigma_x$ and $r_y<N_{\sigma}\sigma_y$ cuts
for the simple single Gaussian shape. The parameter $N_{\sigma}$ is
configurable.  Its default value $N_\sigma=5$ is usually sufficient to reach
the required accuracy. For the single Gaussian bunch the excluded weight,
ie. the density integral of the not simulated region
$(r_x>5\sigma_x \ \text{or}\ r_y>5\sigma_y)$, is
$1-(1-e^{-5^2/2})^2=7.5\cdot10^{-6}$.  To additionally reduce the CPU time,
the largest $(r_x,r_y)$ pairs are not simulated, namely, it is required that
\begin{equation}
\label{eq:88}
\left[\frac{r_x}{n_{max}\Delta_x}\right]^2 + \left[\frac{r_y}{n_{max}\Delta_y}\right]^2 < 1.
\end{equation}
This increases the lost weight to $(5^2/2+1)e^{-5^2/2}=5.0\cdot10^{-5}$.
The remaining weights are normalized.

The default value of the configurable $N_{part}$ parameter from \cref{eq:91}
is $5\,000$. The real number of generated particles is 23\% less due to
\cref{eq:88}.  This default value is used in the simulation examples
discussed in the following.  The other beam parameters are listed in
Table~\ref{tbl1}. They are taken from~\cite{madx12} to compare its old,
biased method used at LHC in 2012--2019 and the results of the B*B
simulation. Increasing $N_{part}$ to $100\,000$ with the default B*B settings 
changes relatively the luminosity integral by $\le 4\cdot10^{-5}$ in the full
simulated range of bunch separations from $0$ to $200\ \mu$m.

\begin{table}[htbp]
  \caption{Bunch parameters from~\cite{madx12} describing one of van der Meer
    calibration scans in ATLAS.}
  \label{tbl1}
  \begin{center}
    \begin{tabular}{c@{\hskip 2mm}c@{\hskip 2mm}c@{\hskip 2mm}c@{\hskip 2mm}c@{\hskip 2mm}c}
    p & $Z_{1,2}$ & $\beta_{x,y}$ & tune $Q_x, Q_y$ & bunch $\sigma$ & $N_{1,2}$\\ \hline
    3500, GeV & 1 & 1.5 m & 64.31, 59.32 & 40 $\mu$m & $8.5\cdot10^{10}$ \\
    \end{tabular}
  \end{center}
\end{table}

\subsection{Stages in the simulation}
The bias associated to the phase space limit \cref{eq:88} and to the
approximation of the continuous integral by the discrete sum in \cref{eq:84}
partially cancels in the ratio
\begin{equation}
\label{eq:90}
R= \frac{\int (\rho_1+\delta\rho_1)\rho_2dx\,dy}{\int \rho_1\rho_2dx\,dy}
\end{equation}
if both integrals are taken numerically in the same way.  Therefore, B*B
starts from simulating $N_{turn}^{no\ BB}$ accelerator turns without the
beam--beam interaction ie. with the beam--beam kick $\Delta u'=0$ in
\cref{eq:81}. The unperturbed luminosity $\int \rho_1\rho_2 dx\,dy$ is
estimated turn-by-turn using \cref{eq:84}. The default value of
$N_{turn}^{no\ BB}$ is conservatively chosen to be 1000. In
Sec.~\ref{sec:lissajous} it will be explained, however, that for the tunes
$Q_{x,y}$ with two digits after the comma, like at LHC, any multiple of 100
leads to identical results. Therefore, $N_{turn}^{no\ BB}=100$ is sufficient
in this case.  With the beam parameters listed in Table~\ref{tbl1} this value
gives the relative deviation between the numerical and analytical
$\int \rho_1\rho_2 dx\,dy$ integrals less than $2\cdot10^{-4}$ in the
practically important region of the beam separations contributing
$\sim 99.9\%$ to the cross-section integral in \cref{eq:vdm}.
The resulting bias of the ratio $R$ in \cref{eq:90} should, therefore, be
smaller.

\begin{figure}[htbp]
  \begin{center}
  \includegraphics[width=0.40\textwidth]{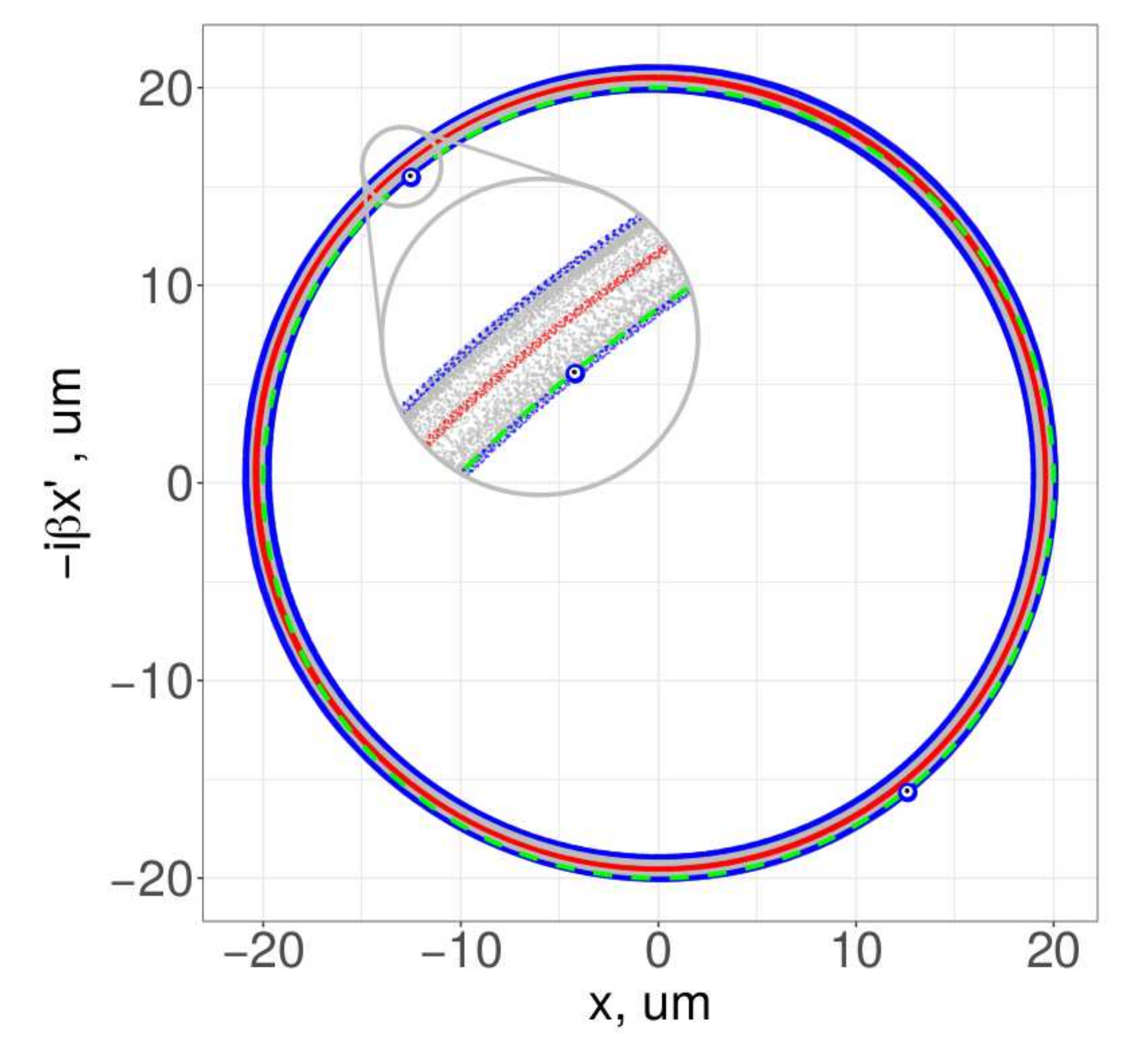}
  \caption{The trajectories of the particles with $r_x=20$, $r_y=30\ \mu$m in
    $z_x$ projection for the bunches separated in $x$ by 40~$\mu$m with the
    parameters from Table~\ref{tbl1} as an example. The outer blue
    trajectories are formed by two initially opposite particles (marked by
    small open circles) after the instantaneous switch of the beam--beam
    force. The grey band between them is composed of such trajectories from
    all particles. The adiabatic trajectory is in red and the initial circle
    with $r_x=20\ \mu$m is shown by the green dashed line.}
  \label{fig4}
\end{center}
\end{figure}

After the first $N_{turn}^{no\ BB}$ turns the beam--beam kick is switched
on. The user has two options: either instantaneously apply the nominal kick
$\Delta u'$ in \cref{eq:81} or increase it slowly or ``adiabatically'',
namely, linearly from zero to the nominal value during $N_{turn}^{adiab}$
turns.  In the former case, the particle trajectory instantaneously changes
from the ideal circle to the one perturbed by the beam--beam force. The
intersection of the two trajectories is the last point on the ideal circle.
It is positioned randomly, and different points on the ideal circle lead to
different perturbed trajectories. This is depicted in Fig.~\ref{fig4}.  Two
initially opposite points, marked in the figure by the small open circles,
create two outer blue trajectories.  Their evolution is followed during $10^7$
turns and every 1000-th point is shown in the plot.  The region in grey in the
middle is filled with all other trajectories. The unperturbed ideal circle
with the center at the origin is shown by the green dashed line.  As one can
see, the center of the blue circles is shifted due to the change of the orbit
in $x$ and $x'$. This will be discussed in more detail in \cref{xchecks}. Note
that the ideal green circle is infinitely ``thin'', but the blue points are
scattered because of the beam--beam $x$-$y$ coupling and the variations in the
other $z_y$-projection. The ``thickness'' of the blue trajectories increases
with the force strength. For much larger forces the trajectory becomes
significantly non-circular.

In the adiabatic case, the trajectories change slowly and the particles have
time to redistribute over them.  The initial position on the circle then has
little importance, and the {\it whole} initial circle transforms to
approximately {\it one} final trajectory shown in red in Fig.~\ref{fig4}.
Here, the beam--beam interaction is slowly switched on during 1000 turns and
then, as in the previous case, every 1000-th turn is shown out of $10^7$ in
total. The red trajectory has approximately the same spread as the blue outer
ones but less than the grey band composed of {\it many} trajectories initiated
by the instantaneous switch.

\begin{figure}[htbp]
  \begin{center}
  \includegraphics[width=0.45\textwidth,angle=0]{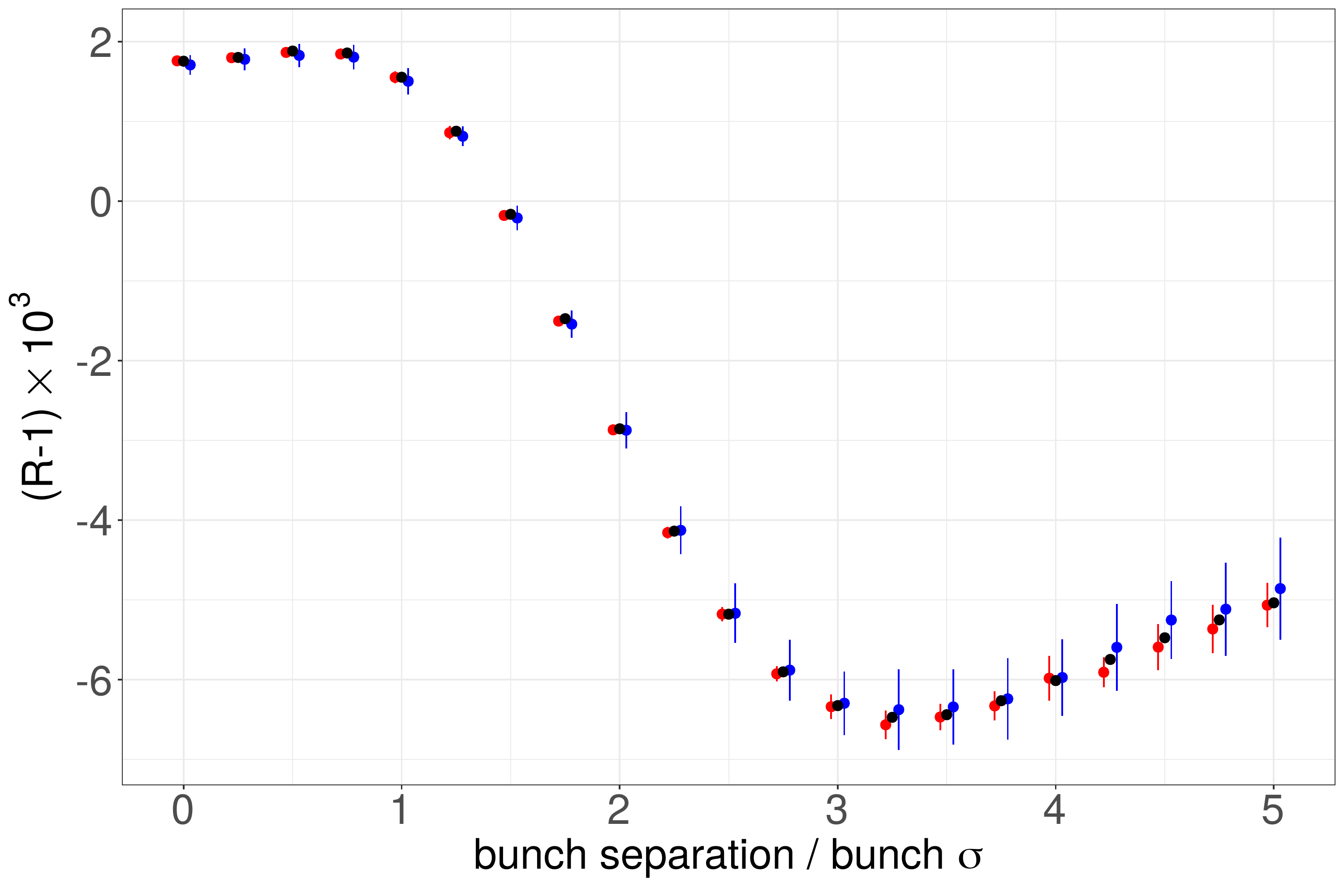}
  \caption{The deviation of the ratio $R$ defined in \cref{eq:90} from unity
    in per mille in the $x$-scan versus the separation of the bunches
    expressed in the bunch widths.  The beam parameters are taken from
    Table~\ref{tbl1}. The error bars show the standard deviation of the
    results when the simulation was repeated with different random seeds 100
    and 25 times for instantaneous (blue) and adiabatic (red points) switch of
    the beam--beam force, respectively. The number of simulated turns are
    $[N_{turn}^{no\ BB}, N_{turn}^{adiab}, N_{turn}^{BB}] = [100,0,500]$
    (blue), $[100,100,500]$ (red), $[10\,000,10\,000,1\,000\,000]$ (black).
    The red and blue points are displaced horizontally to the left and right,
    respectively, to reduce overlapping.}
  \label{fig5}
\end{center}
\end{figure}

In any case, the spread is negligible compared to the width of the opposite
bunch, namely, 40~$\mu$m in Fig.~\ref{fig4}. Therefore, the contribution to
the overlap integral in \cref{eq:84} is essentially determined by the
infinitely ``thin'' {\it average} trajectory, which is the same in both
instantaneous and adiabatic switch cases. The integral does not depend on the
way how the beam--beam force is switched on. This is demonstrated in
Fig.~\ref{fig5} by the blue and red points for various beam separations.

In both cases only 500 (100) accelerator turns were simulated to determine the
perturbed (initial) overlap integral. In the adiabatic case the force was
switched on during 100 turns. The error bars show the standard deviations of
the results obtained with different random generator seeds.  Clearly, the
adiabatic switch is preferable. The instantaneous switch leads to the spread
of the trajectories and larger statistical fluctuations of the final integral.

The central black points in Fig.~\ref{fig5} are the result of the simulation
when the beam--beam force was switched on slowly in $10^4$ turns and the
luminosity integral was calculated over $10^6$ turns. As one can see from
Fig.~\ref{fig5}, simulating only 500 turns already gives
sufficient accuracy. The default B*B values, $N_{turn}^{no\ BB}=1000$,
$N_{turn}^{adiab}=1000$ and $N_{turn}^{BB}=5000$, are, therefore, quite
conservative and may be reduced by a factor of 10 in practice.

After the beam--beam force is fully switched on and before calculating the
overlap integral over $N_{turn}^{BB}$ turns in the final stage, the B*B
simulation has an option to run $N_{turn}^{stab}$ turns for a
``stabilization''. This period is not used for calculating the luminosity
integral. Normally this parameter can be set to zero because, for example, in
the adiabatic case the particle arrives at its final trajectory as soon as the
force reaches its nominal value. The following evolution does not change the
trajectory. $N_{turn}^{stab}$ parameter exists only for flexibility and for
experimenting with the B*B simulation.

\subsection{Lissajous curves}
\label{sec:lissajous}
\begin{figure}[htbp]
  \begin{center}
  \includegraphics[width=0.205\textwidth,angle=0]{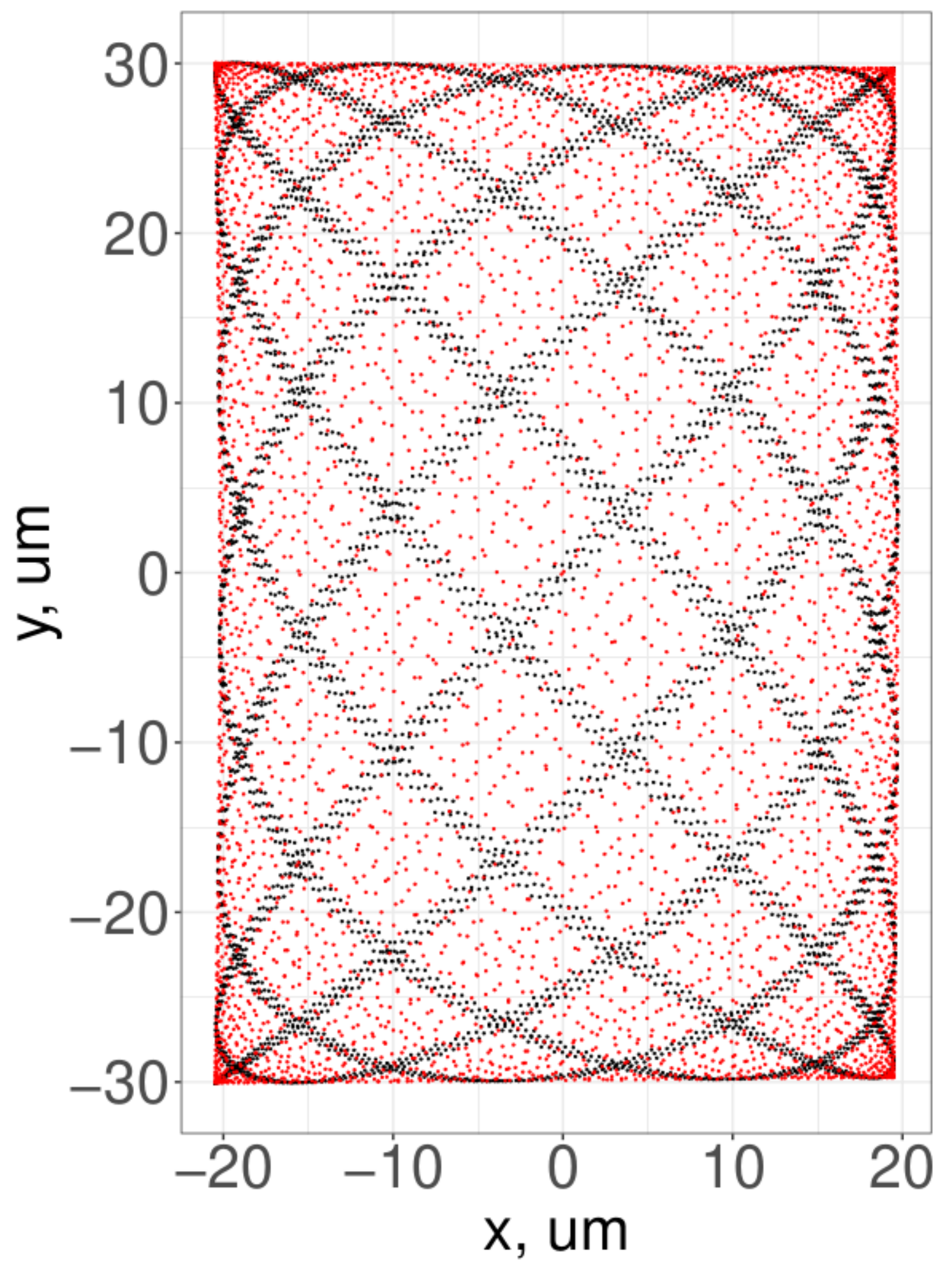}
  \includegraphics[width=0.255\textwidth,angle=0]{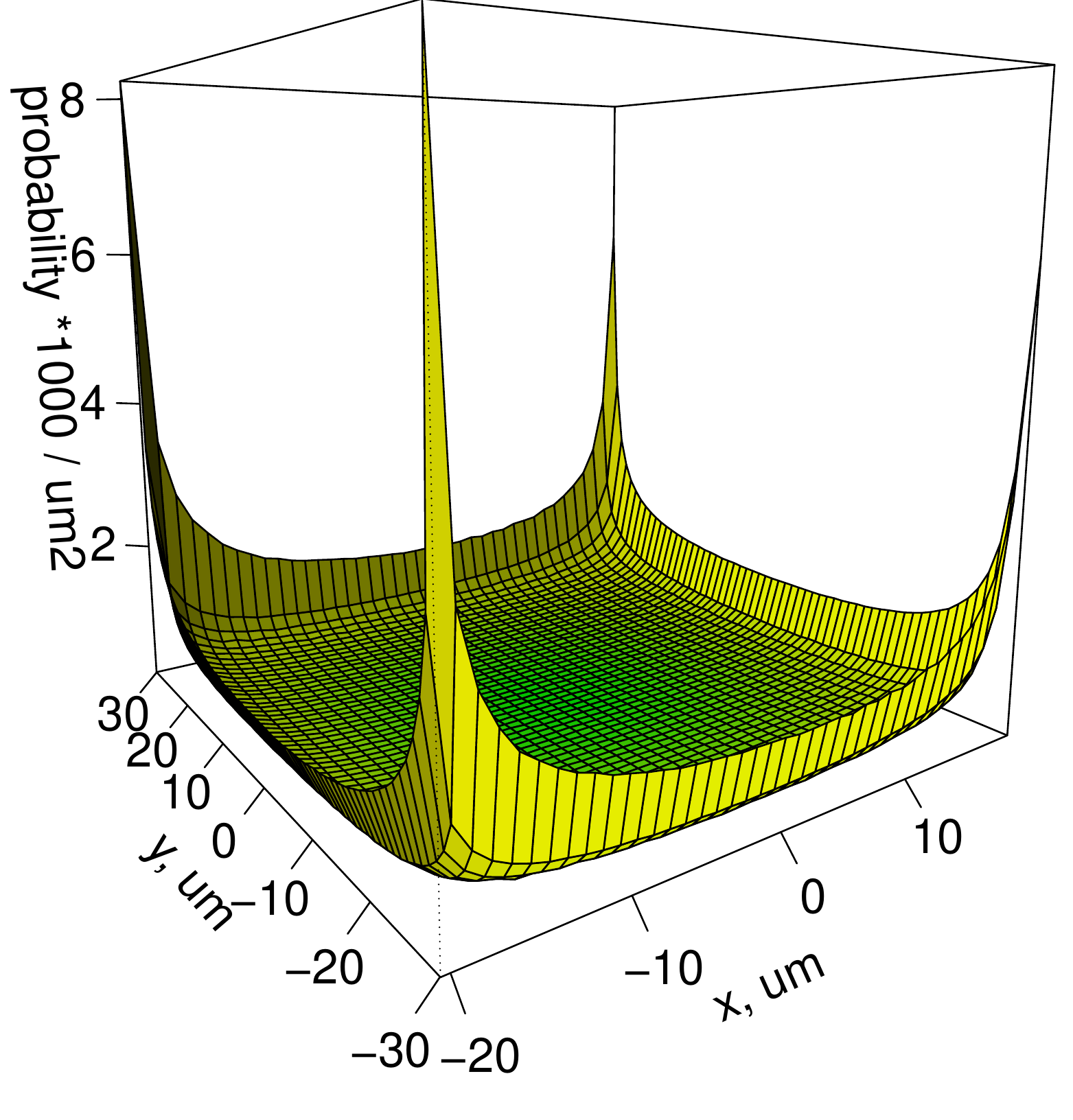}
  \caption{Left: The trajectory of one particle with $r_x=20$, $r_y=30$~$\mu$m
    followed during 5000 turns in the presence of the beam--beam force. The
    bunch parameters are from Table~\ref{tbl1} except the tunes. The black
    points are obtained with $Q_x=0.3$, $Q_y=1/3$. The red points, scattered
    in the background, are simulated with $Q_x$, $Q_y$ shifted by
    $+e^{-8.5}/4, -e^{-8.5}/4$, respectively. Right: the probability density
    of $x$-$y$ localization of the particle followed during $10^6$ turns.}
  \label{fig6}
\end{center}
\end{figure}

Without the beam--beam interaction the $x$-$y$ trajectory of a particle placed
at the $z_x$, $z_y$-circles with the radii $(r_x, r_y)$ and with the initial
phases $\phi^0_{x,y}$ is described by \cref{eq:79}:
\begin{equation}
\label{eq:92}
x_n=r_x \cos(2\pi Q_xn+\phi^0_x), \ \ 
y_n=r_y \cos(2\pi Q_yn+\phi^0_y),
\end{equation}
ie. appears to be a Lissajous curve. The beam--beam force leads to the
dispersion of the trajectory as shown in Fig.~\ref{fig6} left by the black
points for the artificial values $Q_x=0.3$, $Q_y=1/3$ and 5000 accelerator
turns.

In general, for the rational $Q_{x,y}$ and a small beam--beam force the
trajectory ``cycles'' after $LCD(Q_x, Q_y)$ turns where $LCD$ denotes the
lowest common denominator. In the above example $LCD(3/10,1/3) = 30$, so after
one ``cycle'' with 30 turns the phase increments $30\cdot Q_{x,y}$ become
integer and the following turns pass through approximately the same region of
the phase-space. Therefore, they do not immediately improve its sampling.
After many more turns the beam--beam spread forces the points to fill and to
sample the whole rectangle, but then the simulation takes too much CPU time.

To improve the sampling but keep the number of turns low, one can artificially
shift $Q_{x,y}$ by a small amount and, for example, make them irrational. This
``opens'' the Lissajous curve, so that it fills the whole rectangle $|x|<r_x$,
$|y|<r_y$ even without the beam--beam force.  The modification should be
sufficiently small to ensure that the overlap integral changes negligibly. In
B*B two small irrational numbers $\delta Q_{x,y}$ randomly distributed in the
interval $[-\epsilon, \epsilon]$ are added to $Q_x$ and $Q_y$ by default,
where $\epsilon$ is chosen to be $e^{-8.5}/2=1.01734\ldots\times10^{-4}$. As
an example, the points obtained with the shifts $\delta Q_x=\epsilon/2$,
$\delta Q_y=- \epsilon/2$ in 5000 turns are shown in Fig.~\ref{fig6} left by
the red points. They much better fill the rectangle. This modification of the
tunes is configurable and can be switched off if desired.

When the trajectory fills the whole rectangle, the probability density of the
particle peaks at the boundaries. This is shown in Fig.~\ref{fig6} right for
$N_{turn}^{BB}=10^{6}$.  This can be understood from the simple case when the
beam--beam force is absent and at least one of $Q_{x,y}$ is irrational, so that
the Lissajous curve is open. Then, the density is factorizable in $x$ and $y$,
$\rho_{r_x,r_y}(x,y)=\rho_{r_x}(x)\rho_{r_y}(y)$. The terms $\rho_{r_u}(u)$,
where $u=x,y$, can be derived by projecting the uniform density $d\phi/2\pi$ of
the $z_u$-circle with the radius $r_u$ to the $u$-axis:
\begin{equation}
\label{eq:93}
\rho_{r_u}(u) = \frac{1}{2\pi}\left|\frac{d\phi}{du}\right| =
\frac{1}{2\pi r_u\left|\frac{d\cos\phi}{d\phi}\right|} = \frac{1}{2\pi\sqrt{r_u^2-u^2}}.
\end{equation}
Therefore,
\begin{equation}
\label{eq:94}
\rho_{r_x,r_y}(x,y) = \frac{1}{4\pi^2\sqrt{(r_x^2-x^2)(r_y^2-y^2)}}.
\end{equation}

It is interesting that the smooth two-dimensional $x$-$y$ Gaussian shape of
any bunch is always intrinsically composed from such peaking $\rho_{r_x,r_y}$
densities.  As follows from \cref{eq:87}, they should be taken with the
weights $\propto\exp(-r_x^2/2\sigma_x^2-r_y^2/2\sigma_y^2)dr_x^2\,dr_y^2$.
Conversely, only the distributions decomposable to $\rho_{r_x,r_y}$ can
represent the bunch shapes $\rho_{1,2}(x,y)$.

The functions $K_u(s)$ in \cref{eq:51} and the accelerator phase advances
\cref{eq:53} can be modified by configuring the quadrupole currents. The LHC
tune values in all Run 1 and 2 van der Meer scans were deliberately kept
constant: $Q_x=64.31$, $Q_y=59.32$. The fractional parts of these values have
two digits after the comma, and their lowest common denominator is
$LCD(0.31, 0.32)=100$. This is larger than 30 from the example above, so the
phase-space sampling is better.  The tiny tune change $\delta Q \sim 10^{-4}$
mentioned above introduces a sizeable spread to the trajectory only after
$O(10^3)$ turns.  Fig.~\ref{fig5} shows, however, that 500 turns are already
sufficient to get the required accuracy for the bunch settings from
Table~\ref{tbl1} if the beam--beam force is switched on adiabatically. With
500 turns the small irrationality $\delta Q_{x,y}\sim 10^{-4}$ does not
matter. The luminosity bias due to modifications of $Q_{x,y}$ is also
negligible, it is less than $5\cdot10^{-5}$ for all bunch
separations. Therefore, $\delta Q_{x,y}$ plays no role here but might be
useful in the cases when $LCD(Q_x,Q_y)$ is relatively small or to improve the
accuracy of the overlap integrals in experimenting with individual particles.

With a low number of turns $\sim O(100)$, it is important to choose them as
multiples of $LCD(Q_x, Q_y)$. For the LHC tunes, they should be multiples of
100, like $N_{turn}^{no\ BB}=100$, $N_{turn}^{BB}=500$ in Fig.~\ref{fig5}.
This ensures that every particle trajectory is sampled an integer number of
``cycles''.  Not complete ``cycles'' introduce a bias and a dependence on the
initial phases.

\begin{figure}[htbp]
  \begin{center}
  \includegraphics[width=0.45\textwidth,angle=0]{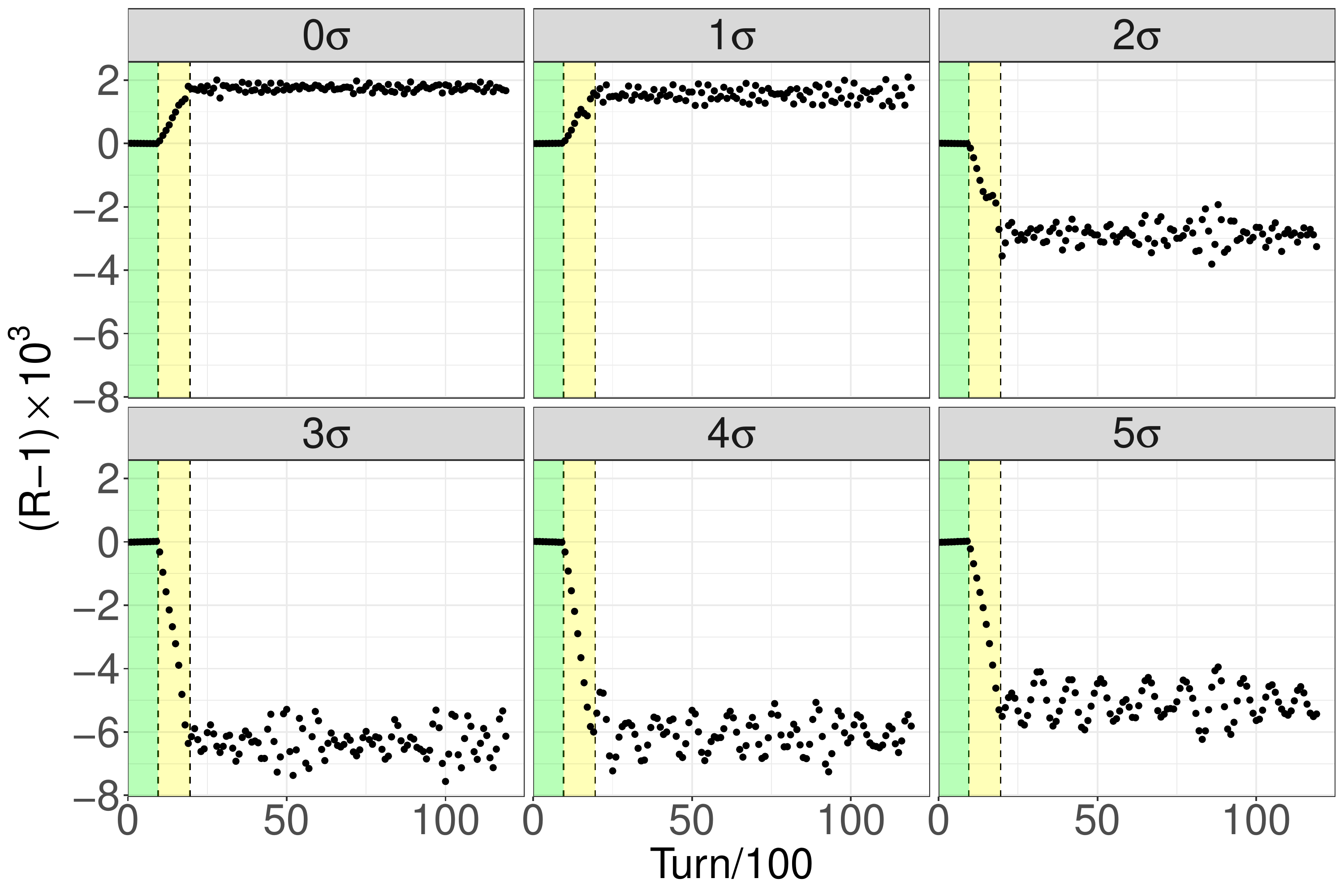}
  \caption{ $(R-1)\times10^3$ in the $x$-scan versus the ``cycle'' number,
    where $R$ is defined in \cref{eq:90} and averaged per ``cycle'',
    ie. within non-intersecting windows of 100 accelerator turns. Green and
    yellow bands delimited by the two vertical dashed lines correspond to
    $N_{turn}^{no\ BB}=1000$ first turns without the beam--beam force and next
    $N_{turn}^{adiab}=1000$ turns when it is being adiabatically switched
    on. The facet labels denote the bunch separation in the bunch sigmas. The
    simulated beam parameters are listed in Table~\ref{tbl1}.}
  \label{fig7}
\end{center}
\end{figure}

Fig.~\ref{fig7} shows the luminosity $R$-correction from \cref{eq:90} in the
$x$-scan calculated ``cycle-by-cycle'', ie.  averaged over non-intersecting
100-turn windows. Six facets correspond to different bunch separations. One
can see small ``oscillations'' for large separations. This is the reason for
larger red error bars in Fig.~\ref{fig5} at higher bins. They show the level
of statistical fluctuations of $R$-averages over $N_{turn}^{BB}=500$ turns,
ie. over the first 5 beam--beam ``cycles'' to the right from the second
vertical dashed line in \cref{fig7}.

\subsection{Beam--beam interactions at several interaction points}

During van der Meer scan, at each LHC experiment there are colliding and not
colliding bunches.  Since ATLAS and CMS detectors are exactly opposite in the
LHC ring, due to this symmetry they share the same colliding bunch pairs. On
the other hand, these pairs never collide in ALICE and LHCb. The latter can
have either ``private'' bunch pairs not colliding anywhere else or pairs with
one or both bunches colliding in one, two or three other experiments. The
beam--beam disturbance at any interaction point propagates everywhere in the
LHC ring and biases van der Meer calibrations in other experiments.

B*B allows to determine the luminosity correction in the general case of one
bunch colliding with an arbitrary number $N$ of fixed bunches of specified
geometries at other interaction points. One should also specify their
beta-functions $\beta^i_{x,y}$ and the constant phase advances normalized by
$2\pi$ 
\begin{equation}
\label{eq:106}
Q^{i}_{x,y}=\frac{1}{2\pi}\int_{s_1}^{s_i} \frac{d\zeta}{\beta_{x,y}(\zeta)}
\end{equation}
between the first and $i$-th points. By definition, the first point has
$Q^1_{x,y}=0$.  The emittance $\epsilon_{x,y}$ of the particle in the linear
accelerator is conserved everywhere in the ring, therefore, the radii of the
circles at different interaction points scale according to \cref{eq:41} as
$r_u^i=\sqrt{\epsilon_u\beta_u^i}\propto \sqrt{\beta_u^i}$. The recurrence
relations then take the form
\begin{align}
\label{eq:95}
  z_{n+1,u}^2 &= \left(z_{n,u}^1 - i\beta_u^1 \Delta u'^1\right)e^{2\pi i\left(Q_u^2-Q_u^1\right)}
                \sqrt{\beta_u^2/\beta_u^1},\\ \nonumber
  z_{n+1,u}^3 &= \left(z_{n,u}^2 - i\beta_u^2 \Delta u'^2\right)e^{2\pi i\left(Q_u^3-Q_u^2\right)}
                \sqrt{\beta_u^3/\beta_u^2},\\ \nonumber
\ldots\\ \nonumber
  z_{n+1,u}^1 &= \left(z_{n,u}^N - i\beta_u^N \Delta u'^N\right)e^{2\pi i\left(Q_u-Q_u^N\right)}
                \sqrt{\beta_u^1/\beta_u^N}.
\end{align}
Let's consider, for example, the last equation. The kick $\Delta u'^N$ is
determined by the electrostatic field at the last interaction point $N$ at the
position $(x^N,y^N)=(Re(z_x^N), Re(z_y^N))$. The factor
$e^{2\pi i\left(Q_u-Q_u^N\right)}$, where $Q_u$ is the full tune, rotates the
phase, while the term $\sqrt{\beta_u^1/\beta_u^N}$ changes the radii scale
from the last to the first interaction point.  Note that to simulate the
perturbation of the second bunch, the order of the interaction points should
be reversed, $N\to (N-1)\to\ldots\to1\to N$, due to its opposite direction.
Also note that, contrary to the overall tunes, the phase advances at LHC are
different for two beams. For example, their values in Run 2 proton--proton van
der Meer scans at $\sqrt{s}=13$~TeV are listed in \cref{tbl2}.

 \begin{table}[htbp]
   \caption{The phase advances $Q_{x,y}^i$ defined in \cref{eq:106} for two
     LHC beams in $x$ and $y$ directions with respect to ATLAS in Run 2
     proton--proton van der Meer scans at $\sqrt{s}=13$~TeV. Last column lists
     the beta-functions $\beta_x=\beta_y$ from~\cite{madx12} used in
     \cref{fig8} simulation.}
   \label{tbl2}
   \begin{center}
     \begin{tabular}
       {c|rr|rr|r}
       &  \multicolumn{1}{c}{$Q_{x}^{\text{beam}\ 1}$} & \multicolumn{1}{c|}{$Q_{x}^{\text{beam}\ 2}$} 
       &  \multicolumn{1}{c}{$Q_{y}^{\text{beam}\ 1}$} & \multicolumn{1}{c|}{$Q_{y}^{\text{beam}\ 2}$}
       & $\beta_{x,y}$, m \\ \hline
        ATLAS&       0 &      0 &      0 &      0 & 1.5\\
        ALICE&  8.2960 & 8.2728 & 7.6692 & 7.9577 & 10 \\
        CMS  & 31.9757 &31.9844 &29.6486 &29.7613 & 1.5\\
        LHCb & 56.0648 &55.7990 &51.0171 &51.7158 & 3  \\
     \end{tabular}
   \end{center}
 \end{table}

\begin{figure}[htbp]
  \begin{center}
  \includegraphics[width=0.49\textwidth,angle=0]{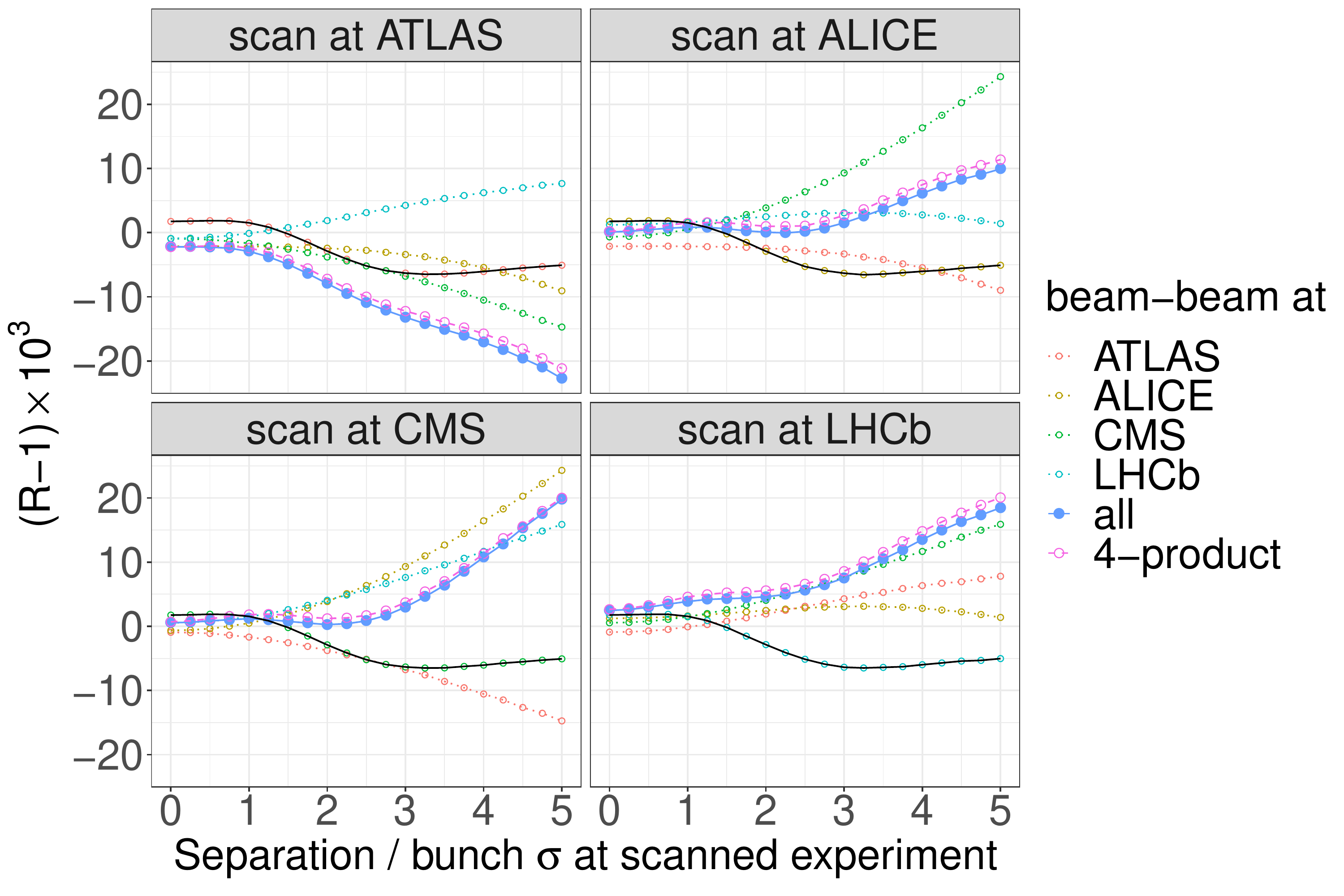}
  \caption{$(R-1)\times10^3$ versus the beam separation expressed in bunch
    widths.  The facets correspond to the $x$-scans at four LHC interaction
    points $i=1,2,3,4$. The solid black (colored dotted) lines with the
    smaller open circles correspond to the correction $R_{i,i}$ ($R_{i,j}$,
    $j\ne i$) when the beam--beam force is switched on only at the scanned LHC
    interaction point $i$ (one of three other points $j\ne i$). The case when
    the beam--beam force is switched on at all four points, $R_{i,1234}$, is
    shown by the blue solid curves with the larger solid circles. The purple
    dashed lines with the larger open circles are their approximations by the
    four-products $R_{i,1}\cdot R_{i,2}\cdot R_{i,3}\cdot R_{i,4}$.  The beams
    at not scanned interaction points collide head-on. The simulation
    parameters are listed in \cref{tbl1,tbl2}.}
  \label{fig8}
\end{center}
\end{figure}

 \cref{fig8} shows how the luminosity in a scan at one LHC interaction point
 is affected by the beam--beam force at other points. As an example, here
 the bunch of the {\it first} beam colliding {\it at all four} LHC experiments
 is simulated. The luminosity change due to the beam--beam perturbation of
 only this bunch is presented in the plot.  To get the full change, a similar
 contribution due to the perturbation of the opposite bunch should be
 simulated and added. Note that each of the four opposite bunches might
 collide with other bunches at other interaction points and this needs to be
 included in its simulation.

The $\beta$-function values for \cref{fig8} simulation are listed in
\cref{tbl2}.  Together with the ATLAS beam parameters from \cref{tbl1} they
are taken from~\cite{madx12}. The number of particles and emittances of all
bunches are the same. The emittances are determined from the bunch width and
$\beta$-value at ATLAS.  Four plots in Fig.~\ref{fig8} show the
$R$-corrections from \cref{eq:90} when the scan is performed along the
$x$-axis at ATLAS, ALICE, CMS and LHCb, respectively.  Let's denote the
scanned experiment by the index $i=1,2,3,4$. The corrections are calculated
and color-coded separately when the force is switched on only at the $j$-th
LHC interaction point ($R_{i,j}$) or simultaneously at all ($R_{i,1234}$). One
can see that the latter, shown by the solid circles, is relatively close to the
product of the four corrections $R_{i,1}R_{i,2}R_{i,3}R_{i,4}$ each calculated
in the absence of the beam--beam force at three other points.

In addition to the luminosity corrections at the scanned interaction point,
B*B automatically calculates similar corrections at all other experiments
where the beams remain stationary. In the simulation of \cref{fig8}, they
collide head-on.  The corresponding $R$-corrections are not shown because they
are not needed for van der Meer calibration.  They less depend on the
separation and are closer to unity, the mismatches do not exceed
$4\cdot10^{-3}$. B*B also allows to simulate the beam--beam interactions at
multiple points and to calculate all luminosity corrections when the scans are
performed simultaneously in several experiments.

\subsection{Invariance under $\beta_{x,y}$, $p$, $\epsilon$ and $N$ scalings}
\label{sec:inv} 
As one can see from Table~\ref{tbl2}, the beta-functions $\beta^i_{x,y}$ and,
therefore, the bunch widths $\sigma_i\propto \sqrt{\beta^i}$ in the simulated
example are different at ATLAS/CMS, ALICE and LHCb. ALICE has the largest
bunches and the smallest beam--beam kick.

The solid black curves in Fig.~\ref{fig8} correspond to $R_{i,i}$ correction
when the beam--beam force is switched on only at the {\it scanned} experiment.
One can see that all such curves in the four facets are the same and also
identical to Fig.~\ref{fig5}.  Let's understand why this happens despite
different bunch widths and kicks.

Scaling of both $\beta_x$ and $\beta_y$ by a constant factor
$\beta_{x,y}\to\alpha\beta_{x,y}$ changes the linear scale in the $z_{x,y}$
planes by $\sqrt{\alpha}$. The two-dimensional electrostatic field from a
point charge $q$ drops with the distance as $\propto 1/R$, so the angular kick
changes as $\Delta u'\to\Delta u'/\sqrt{\alpha}$. According to the definition
\cref{eq:78}, in the $z_u$ complex planes the angular kick is additionally
multiplied by $\beta$, so $\Delta z_u = -i\beta_u\Delta u'$ scales as
$\Delta z_u\to \sqrt{\alpha}\Delta z_u$.  Therefore, the distributions of two
bunches {\it and} the kicks $\Delta z_{x,y}$ scale proportionally to
$\sqrt{\alpha}$.  New particle trajectories can be obtained by the simple
$\sqrt{\alpha}$-scaling of the whole $z_{x,y}$ planes. This modifies the
luminosities by $\alpha$ but keeps constant their ratios, eg. $R$ in
\cref{eq:90}.

To stress this invariance, it is better to rewrite \cref{eq:95} via the ratios
$z_u/\sqrt{\beta_u}$:
\begin{equation}
\label{eq:96}
\frac{z_{n+1,u}^{"i+1"}}{\sqrt{\beta_u^{"i+1"}}} =
\left(\frac{z_{n,u}^i - i\beta_u^i \Delta u'^i}{\sqrt{\beta_u^i}}\right)e^{2\pi i\left(Q_u^{"i+1"}-Q_u^i\right)},
\end{equation}
where $"i+1"$ index denotes the next interaction point, eg. the first after
the last, while $Q_u^{"N+1"}$ is the full tune $Q_u$.

Variables in the complex planes $z_u/\sqrt{\beta_u}$ have another advantage:
they do not change across the interaction points in the ideal linear
accelerator. Indeed, both the phase $Arg(z_u)=\phi_u$ and the absolute value
$\left|z_u/\sqrt{\beta_u}\right|=\sqrt{\epsilon_{u}}$ are invariant, since the
emittance $\epsilon_u$ is conserved. So, the complex planes
$z_x/\sqrt{\beta_x}$ and $z_y/\sqrt{\beta_y}$ are common to all interaction
points.  As we have seen, the beam--beam kick
$- i\beta_u^i \Delta u'^i/\sqrt{\beta_u^i}$ in \cref{eq:96} happen to be
invariant to the scaling $\beta_{x,y}^i\to\alpha^i\beta_{x,y}^i$ with
arbitrary $\alpha_i$ because of $\propto 1/R$ Coulomb's law in two
dimensions. Therefore, the trajectories in the $z_u/\sqrt{\beta_u}$ planes
drawn from the fixed initial points $z_{u}^1/\sqrt{\beta_u}$ using
\cref{eq:96} do not depend at all on $\beta_u^i$ values.  This is true under
the condition that $\beta_x^i$ and $\beta_y^i$ {\it scale proportionally} but
independently in different experiments. If $\beta_x^i/\beta_y^i$ ratio changes
at any point $i$, this also changes the relative $x^i$ and $y^i$ scales in
$R^i=\sqrt{(x^i)^2+(y^i)^2}$ and in the Coulomb's law, and breaks the kick
invariance.  At LHC, however, typically at all interaction points the two
beta-functions are equal, $\beta_x^i=\beta_y^i$.

The horizontal axes in \cref{fig5,fig8} are also chosen in the form of the
scale-invariant ratio. Therefore, all $R_{i,i}$-correction curves shown in
these figures do not depend on the specific $\beta_x^i=\beta_y^i$ values from
\cref{tbl1,tbl2}. They are determined only by the distribution of the {\it
  emittances} $\epsilon_{u}$ or the areas of the $z_{u}$-circles
$\pi r_u^2=\pi\epsilon_{u}$. For example, $\beta^i=1$~m at all four
experiments leads to the same figures.

Contrary to the tune values, the phase advances are specific for each
experiment. Therefore, with multiple interaction points this symmetry between
the experiments is lost and $R_{i,1234}$ and $R_{i,j}$, $i\ne j$ curves in
\cref{fig8} are different.

There is another interesting invariance of the beam--beam
perturbations. During acceleration of the particles, their angles $u'$
decrease due to the growth of the longitudinal momentum. If the acceleration
is sufficiently slow, like at LHC, the emittance decreases according to the
``adiabatic damping'' formula
\begin{equation}
\label{eq:113}
\epsilon_{u} = \tilde\epsilon_{u}\cdot m/ p,
\end{equation}
where $m$ is the mass of the particle and $\tilde\epsilon_{u}$ is the constant
{\it normalized emittance}. Therefore, when $p$ increases, the complex plane
$z_{u}/\sqrt{\beta_{u}}$ shrinks as $1/\sqrt{p}$ and the beam--beam force
increases as $\sqrt{p}$. The {\it angular} beam--beam kick $\Delta p/p$
contains $p$ in the denominator as in \cref{eq:8}, and decreases as
$1/\sqrt{p}$, ie. coherently with the $z_{u}/\sqrt{\beta_{u}}$ complex
plane. Therefore, in the complex plane corresponding to the {\it normalized}
emittances $z_{u}\sqrt{p/m\beta_{u}}= \sqrt{\tilde\epsilon_{u}}e^{i\phi_u}$
the trajectories remain invariant if the phase advances are constant and
$\tilde\epsilon_{u}$ are conserved. The beam--beam perturbations of the bunches
with identically distributed normalized emittances are identical.

Finally, let's formulate the scaling law for the number of particles $N_2$ in
the opposite bunch. It will be called ``second'' in the following to
distinguish from the ``first'' bunch with the studied trajectories.  The bunch
variables will be marked by the corresponding indices.

Multiplying the normalized emittance of the first bunch by $\alpha$ increases
the linear scale by $\sqrt{\alpha}$ and reduces the beam--beam angular kick by
$1/\sqrt{\alpha}$. In order to have the same $\sqrt{\alpha}$ change in the
linear and angular scales for the first bunch, the beam--beam force from the
second should be enhanced by $\alpha$.  This can be achieved by increasing
$Z_1Z_2N_2$. Therefore, the simultaneous scaling of $\tilde\epsilon_{1u}$ and
$Z_1Z_2N_2$ by the same factor changes the scale but not the {\it shapes} of
the first bunch trajectories. It also keeps constant the luminosity correction
ratios.  If the trajectories are analyzed in the complex plane
\begin{equation}
\label{eq:14}
z_{1u}\sqrt{\frac{p_1}{m_1\beta_{u1}Z_1Z_2N_2}}=
\sqrt{\frac{\tilde\epsilon_{1u}}{Z_1Z_2N_2}}e^{i\phi_{1u}},
\end{equation}
the results
depend only on the initial distributions, the phase advances and
$\beta_{1x}/\beta_{1y}$ ratios, but do not depend on the individual values of
$\beta_{1x,1y}$, $\epsilon_{1u}$, $p_1$, $Z_{1,2}$ or $N_2$. Only their
combination
\begin{equation}
\label{eq:122}
\frac{\tilde\epsilon_{1u}}{Z_1Z_2N_2} = \frac{\epsilon_{1u}p_1}{Z_1Z_2N_2m_1}
\end{equation}
matters. This can also be confirmed with the B*B simulation.

\subsection{Simulation of the beam crossing angle}
If the beams collide with the crossing angle as in \cref{fig1}, the betatron
oscillations occur in the planes {\it transverse to the
  beam vectors} $\vec{v}_{1,2}$. They determine the transverse widths
$\sigma_{iT}$ of the bunches $i=1,2$.  The luminosity, however, depends on the
bunch widths $\sigma_{iT}'$ {\it perpendicular to} $\Delta \vec{v}$. According to
\cref{eq:39}, they get additional contributions from the longitudinal widths
$\sigma_{iL}$.

The beam--beam kick is also {\it perpendicular to} $\Delta \vec{v}$ either in
the laboratory or, for example, in the rest frame of the first bunch particle
$q_1$ shown in \cref{fig2}.  Therefore, $q_1$ ``sees'' the opposite bunch
width $\sigma_{2T}'$ {\it enhanced} by $\sigma_{2L}$, and this value should be used
in the B*B simulation for $\rho_2$. The bunch creating the field is ``static''
in the B*B model, its betatron transverse motion and $\sigma_{2T}$ do not
matter.

However, to simulate the betatron trajectories in the first bunch, the initial
widths $\sigma_{1x}$, $\sigma_{1y}$ should be specified in B*B {\it without} the
longitudinal component. The betatron oscillations are insensitive to
$\sigma_{1L}$ spread.  Since the crossing angles are small at LHC,
$\alpha_{1,2}=\alpha< 10^{-3}$, the kicks calculated in the primed frame with
the parallel beams can be propagated without changes to the frame of the
betatron motion.  Similarly, the simulated $x$, $y$ beam transverse
coordinates can be propagated back for the luminosity calculation.

In this way the luminosity correction is determined for the unperturbed widths
$\sigma_{1T}, \sigma_{2T}'$ instead of the required
$\sigma_{1T}', \sigma_{2T}'$. Therefore, the final density
$\rho_1+\delta\rho_1$, corresponding to $\sigma_{1T}$, should be additionally
smeared by the contribution from $\sigma_{1L}$. This is performed in the B*B
simulation in the following way.  Let's consider the general case when neither
$x$ nor $y$ of the first bunch lies in the crossing plane, and denote by
$\beta_1$ the angle between the $x$-axis and the projection of the crossing
plane to the $x$-$y$ plane. The perturbed density $\rho_1+\delta\rho_1$ should
then be convolved with the two-dimensional Gaussian $G(x,y)$ with the sigmas
$\alpha\sigma_{1L}\cos\beta_1$ and $\alpha\sigma_{1L}\sin\beta_1$ in $x$
and $y$, respectively. If, for example, the $x$-axis belongs to the crossing
plane and $\beta_1=0$, the smearing occurs only along $x$ with the sigma
$\alpha\sigma_{1L}$, while $G(x,y)$ projection to $y$ reduces to the
delta-function. Instead of smearing $\rho_1+\delta\rho_1$ in the overlap
integral, it is simpler to perform an equivalent smearing of $\rho_2$:
\begin{align}
\label{eq:115}
  L &\propto \int((\rho_1+\delta\rho_1)*G)\,\rho_2\,d^2\vec{r}\nonumber\\
  &=\int\left(\int(\rho_1+\delta\rho_1)(\vec{r}')G(\vec{r}-\vec{r}')d^2\vec{r}'\right)
      \rho_2(\vec{r})d^2\vec{r}\nonumber\\
    &=\int (\rho_1+\delta\rho_1)(\vec{r}')\left(\int G(\vec{r}'-\vec{r})
      \rho_2(\vec{r})d^2\vec{r}\right)d^2\vec{r}'\nonumber\\
  &= \int(\rho_1+\delta\rho_1)\,(G*\rho_2)\,d^2\vec{r}',
\end{align}
since $G(\vec{r}-\vec{r}')=G(\vec{r}'-\vec{r})$, where $\vec{r}$ and
$\vec{r}'$ denote the two-dimensional vectors in the $(x,y)$ plane.

To implement this scheme, the B*B simulation takes the values
\begin{equation}
\label{eq:11}
\Delta\sigma_{1x}^i = \alpha\sigma_{1L}\cos\beta_1,\quad \Delta\sigma_{1y}^i = \alpha\sigma_{1L}\sin\beta_1
\end{equation}
as configurable parameters defining the Gaussian $G^i(x,y)$ for each
interaction point $i$. Its convolution with the multi-Gaussian $\rho_2$ is
performed analytically. The result is used in \cref{eq:84} instead of $\rho_2$
for the overlap integral calculation. Note that it receives contributions from
both $\sigma_{1L}$ and $\sigma_{2L}$.  The field is determined from $\rho_2$
not smeared by $G^i(x,y)$, where only $\sigma_{2L}$ contributes. The beam--beam
perturbation of the longitudinal beam dynamics is neglected and the
longitudinal spread is always approximated by a single Gaussian. By default
the parameters \cref{eq:11} are zero.

\subsection{Cross-checks and comparison with the old model used at LHC in
  2012--2019}
\label{xchecks}
In \Cref{avrkick} it was shown that the full beam--beam force exerted on the
first bunch can be easily calculated as the force between the first bunch
collapsed to the origin $(0,0)$ and the second one with the density
``inflated'' from $\rho_2$ to $\tilde\rho_1*\rho_2$. In case of the Gaussian
bunches with the widths $\sigma_{1,2;x,y}$ and the centers $\vec{r}^0_{1,2}$,
the cross-correlation $\tilde\rho_1*\rho_2$ is also Gaussian and has the
widths $\Sigma_{x,y}=\sqrt{\sigma_{1;x,y}^2+\sigma_{2;x,y}^2}$ and the center
$\Delta\vec{r}_{21}^0=\vec{r}^0_2-\vec{r}^0_1$. Its electrostatic field $E_{2u}$, given
by \cref{eq:2} or~\cref{eq:BassettiErskine}, allows to calculate the {\it
  average angular} beam--beam kick $\Delta u'=q_1E_{2u}/p_1c=eZ_1E_{2u}/p_1c$,
of the first bunch particles. This value is {\it exact} if all bunch particles
move with the opposite and constant velocities close to the speed of light, so
that one can calculate the kicks using the laws of electrostatics.

Let's substitute the beam--beam angular kick by its average. The $x$- and
$y$-directions then decouple. To simplify notations, let's drop the coordinate
subscript $u$, denote the one-turn phase advance by $2\pi Q=\phi$, introduce
the constant $\Delta=\beta_1\Delta u'=\beta_1q_1E_{2u}/p_1c$ and consider for
simplicity the case with only one interaction point. The recurrence relation
\cref{eq:81} then defines the geometric sequence
\begin{align}
\label{eq:9}
  z_{n+1} &= (z_n - i\Delta)e^{i\phi} \nonumber\\
          &= z_1e^{in\phi} -i\Delta e^{i\phi}\left(1+e^{i\phi}+e^{2i\phi}+\ldots+e^{i(n-1)\phi}\right) \nonumber\\
  &=z_1e^{in\phi} -i\Delta e^{i\phi}\frac{e^{in\phi}-1}{e^{i\phi}-1}.
\end{align}
This can be rewritten as
\begin{equation}
\label{eq:98}
  z_{n+1}-\frac{i\Delta e^{i\phi}}{e^{i\phi}-1}=\left(z_1 -\frac{i\Delta e^{i\phi}}{e^{i\phi}-1}\right)e^{in\phi}.
\end{equation}
A comparison with the equation $z_{n+1}=z_1e^{in\phi}$ without the beam--beam
interaction shows that the constant kick only shifts the center of rotation
from the origin to the point
\begin{equation}
\label{eq:104}
z_0=\frac{i\Delta e^{i\phi}}{e^{i\phi}-1} = \frac{
  e^{i\phi/2}}{2\sin \frac{\phi}{2}}=
\frac{\Delta}{2\tan\frac{\phi}{2}}+i\frac{\Delta}{2}.
\end{equation}
Its real part determines the shift of the beam orbit, while the scaled
imaginary part $-Im(z_0)/\beta_1=-\Delta u'/2$ gives the angular shift just
{\it before} the beam--beam interaction. {\it After} receiving the kick
$+\Delta u'$, it flips from $-\Delta u'/2$ to $+\Delta u'/2$, while the next
accelerator turn changes it back to $-\Delta u'/2$.

The orbit shift expressed in the bunch widths should be invariant and should
depend only on $\tilde\epsilon_1/Z_1Z_2N_2$. Indeed, since
$\Delta=\beta_1\Delta u'=\left(\sigma_1^2p_1/\tilde\epsilon_1 m_1\right)\cdot\left(eZ_1E_{2u}/p_1c\right)$, it can
be written as
\begin{equation}
\label{eq:119}
\frac{Re(z_0)}{\sigma_1}=
\frac{e(E_{2u}\sigma_1)}{Z_2N_2}\cdot\frac{1}{2m_1c\tan \frac{\phi}{2}}\cdot \frac{Z_1Z_2N_2}{\tilde\epsilon_1}.
\end{equation}
Here, $E_{2u}$ has $\propto Z_2N_2/R$ dependence, so
$e(E_{2u}\sigma_1)/Z_2N_2$ is invariant. For example, for the round bunches
with $\sigma_1=\sigma_2=\sigma$, $\Sigma_{x,y}=\sqrt{2}\sigma$ separated in
$x$ by $\Delta x$, substituting $E_{2u}$ from~\cref{eq:2} or \cref{eq:13}
gives
\begin{equation}
\label{eq:120}
\frac{Re(z_0)}{\sigma}=
\frac{1-e^{-\Delta x^2/4\sigma^2}}{\Delta x/\sigma}\cdot\frac{\alpha\hslash}{m_1\tan \frac{\phi}{2}}\cdot
\frac{Z_1Z_2N_2}{\tilde\epsilon_1}.
\end{equation}
The maximal value of the first term depending only on $\Delta x/\sigma$ is
$0.31908\ldots\ $. Therefore, for protons one has
\begin{equation}
  \label{eq:121}
  \frac{Re(z_0)}{\sigma}\le
  \frac{N_2}{(2.0420\ldots \times 10^{12})\cdot[\tilde\epsilon_1/\mu\mathrm{m}]\cdot\tan\frac{\phi}{2}},
\end{equation}
which is less than 1\% for the typical LHC values
$\tilde\epsilon_1\approx 3-4\ \mu$m, $N_2\approx (7-9)\cdot10^{10}$ and
$\phi_x/2=0.31\pi$, $\phi_y/2=0.32\pi$.

In case of several interaction points with the phase advances
$2\pi Q_i=\phi_i$ with respect to the first point where by definition
$\phi_1=0$, and the constant kicks $\Delta_i$, $i=1,2,\ldots,N$, the first
equation in \cref{eq:9} modifies to
\begin{equation}
\label{eq:107}
  z_{n+1} =z_ne^{i\phi}-i\Delta_1e^{i\phi}-i\Delta_2e^{i(\phi-\phi_2)}-\ldots-i\Delta_Ne^{i(\phi-\phi_N)},
\end{equation}
which becomes equivalent to \cref{eq:9} after the substitution
\begin{equation}
\label{eq:108}
\sum_{i=1}^N\Delta_ie^{-i\phi_i}=\Delta.
\end{equation}
So, we have again a circular trajectory with the center shifted to
\begin{align}
\label{eq:109}
z_0&=\frac{i\left(\sum_{i=1}^N\Delta_ie^{-i\phi_i}\right)e^{i\phi}}{e^{i\phi}-1}=
\frac{\sum_{i=1}^N\Delta_ie^{i[\phi/2-\phi_i]}}{2\sin \frac{\phi}{2}}\nonumber\\
   &= \frac{\sum_{i=1}^N\Delta_i\cos(\frac{\phi}{2}-\phi_i)}{2\sin \frac{\phi}{2}}+
     i\frac{\sum_{i=1}^N\Delta_i\sin(\frac{\phi}{2}-\phi_i)}{2\sin \frac{\phi}{2}},
\end{align}
whose real and opposite imaginary parts define the spatial and
$\beta_1$-scaled angular shifts, respectively. This formula is very well known
in the accelerator physics, which uses different tools for its
derivation~\cite{Chao:1490001}.

For experimenting and debugging, B*B has configurable options to substitute
the exact kick formula by its average and to output the $x$, $y$ orbit
shifts. The latter are calculated as weighted sums $\sum_{i=1}^N w_iu_i$ over
the simulated particle coordinates $u_i$, averaged over accelerator turns
similarly to \cref{eq:84}.  For example, with
$(N_{turn}^{no\ bb},N_{turn}^{adiab},N_{turn}^{bb})=(100,100,500)$, default
other settings and the single interaction point with the parameters from
\cref{tbl1}, the mismatch between the simulated and predicted orbit shifts
averaged over the beam separation range $[0-200]\ \mu$m has the standard
deviation $\sigma\lesssim 2$~nm. The maximal predicted orbit shift is
$0.28\ \mu$m. 
The simulated shifts without the beam--beam force are compatible with zero
with $\sigma\lesssim 0.3$~nm.

If the beam--beam interaction is small, even with the {\it exact}, not constant
kick the particle trajectory remains approximately circular as illustrated in
\cref{fig4}. One particle trajectory corresponds to two circles in
$z_{x,y}$-planes.  The orbit shift is then {\it approximately} determined by
the beam--beam force $\vec{F}^{tr}$ averaged over the trajectory like shown in
\cref{fig6}, with the $x$-$y$ density projection \cref{eq:94}.  The circle from
\cref{fig4} in the $z_x$-plane shifts according to the average $x$-component
of the force $F^{tr}_x$, dependent, however, on the radius in $z_y$-plane. So,
different $z_{x,y}$-circle pairs have different $\vec{F}^{tr}$ and shift
differently. However, the ratio $r_{Re/Im}$ of the real and imaginary parts in
equations \cref{eq:104,eq:109} remains {\it constant} regardless of
$\vec{F}^{tr}$. Therefore, the shifted centers $z_0=x_0+iy_0$ lie on the line
$r_{Re/Im}=x_0/y_0$ and the spatial and angular shifts are {\it
  proportional}. Averaging over the bunch should preserve this
proportionality.  Therefore, if $\Delta$, $\Delta_i$ denote {\it exactly}
known average bunch angular kicks in \cref{eq:104,eq:109}, these equations
should correctly predict the {\it average} $z_0$ shifts. For
example, the real part $Re(z_0)$ predicts the spatial average or {\it ``orbit''}
shifts. Note that contrary to the exact {\it average angular kicks}
$\Delta u'$ and $\Delta$, the values of $z_0$, ie. the {\it angular} and the {\it
  orbit shifts} calculated from \cref{eq:104,eq:109}, depend on the assumption
that the trajectories are approximately circular, which is violated for strong
kicks or a large number of interaction points.

\begin{figure}[htbp]
  \begin{center}
  \includegraphics[width=0.49\textwidth,angle=0]{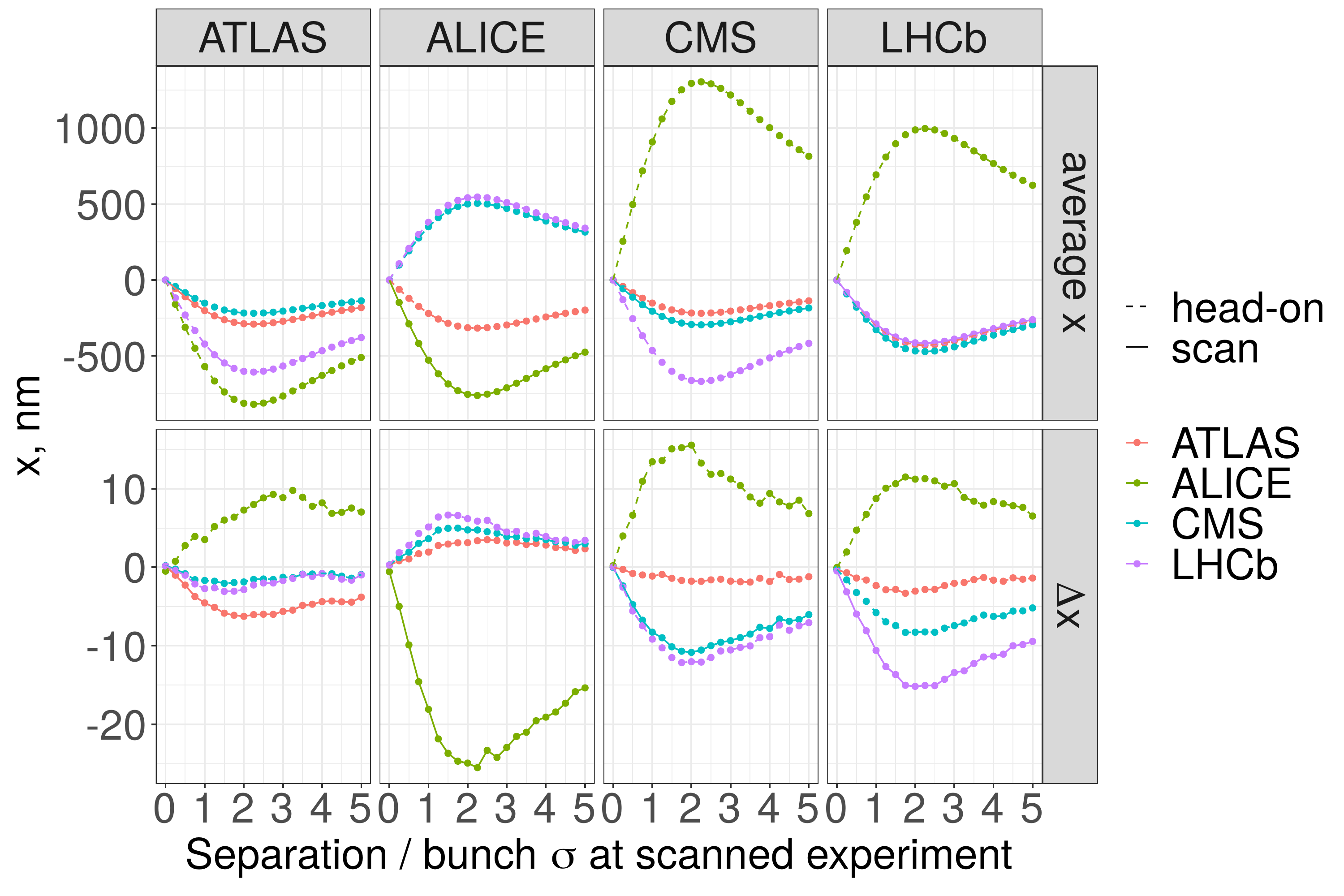}
  \caption{Upper row: $x$-coordinate averaged over the first bunch at four
    experiments, lower row: its deviation from \cref{eq:109} approximation,
    both in nm, versus the beam separation expressed in the bunch widths. The
    column labels denote the scanned interaction points (marked by the solid
    lines in the plots), at other points the beams collide head-on (the dashed
    lines).  The simulation is the same as in \cref{fig8}.}
  \label{fig9}
\end{center}
\end{figure}

As an example, \cref{fig9} shows the $x$-orbit shifts in the same simulation
as in \cref{fig8}.  The beam--beam kick is switched on at all four interaction
points.  The mismatches between the shifts calculated by the B*B simulation
and the analytic approximation \cref{eq:109} reach 26~nm. Although small in
the absolute scale, they are significantly larger than the statistical
fluctuations. The orbit shift along $y$-coordinate, where the beams are not
separated, is compatible with zero within $\sigma=0.4$~nm.

\vspace{5mm}

In the old beam--beam model used at LHC in 2012--2019, the field $\vec{E}_2$
in the momentum kick $\Delta\vec{p}_1=q_1\vec{E}_2/c$ was not taken from
\cref{eq:2} or \cref{eq:BassettiErskine} but was approximated by a simple
linear function of the $x$ and $y$ coordinates. The constant term was chosen
to reproduce the orbit shift from \cref{eq:104}. The slopes were taken from
the derivatives $\partial\Delta\vec{p}_1/\partial x$,
$\partial\Delta\vec{p}_1/\partial y$ at the first {\it bunch center.} The
model was limited to the case when the Gaussian bunches collided {\it head-on
  in the not scanned coordinate.}  Under this assumption, the
cross-derivatives $\partial\Delta p_{1,y}/\partial x$,
$\partial\Delta p_{1,x}/\partial y$ vanish, and the $x$, $y$-slopes were taken
as $\partial\Delta p_{1,x}/\partial x$, $\partial\Delta p_{1,y}/\partial y$,
respectively. They can be calculated from \cref{eq:2} for the round
bunches. For example, in the $x$-scan at the center of the first bunch, where
$R=\Delta x$, one has
\begin{align}
\label{eq:110}
\left.\frac{1}{k_1}\frac{\partial\Delta p_{1,x}}{\partial x}\right|_{x=\Delta x,y=0} &= 
  -\frac{1}{\Delta x^2}+
  \left(\frac{1}{\sigma_2^2}+\frac{1}{\Delta x^2}\right) e^{-\Delta x^2/2\sigma_2^2},\nonumber\\
 \left.\frac{1}{k_1}\frac{\partial\Delta p_{1,y}}{\partial y}\right|_{x=\Delta x,y=0} &=
                                              \frac{1 - e^{-\Delta x^2/2\sigma_2^2}}{\Delta x^2},
\end{align}
where $k_1=2\alpha\hslash Z_1Z_2N_2$.  The full kick at $\Delta x$-separation was
approximated as the following linear function of $x$ and $y$:
\begin{align}
\label{eq:111}
  \Delta p_{1,x}& \approx
                  x\cdot \left.\frac{\partial\Delta p_{1,x}}{\partial x}\right|_{x=\Delta x,y=0} +
                   k_1\frac{1-e^{-\Delta x^2/2(\sigma_1^2+ \sigma_2^2)}}{\Delta x}
                              \nonumber\\
  \Delta p_{1,y}& \approx 
                              y\cdot\left.\frac{\partial\Delta p_{1,y}}{\partial y}\right|_{x=\Delta x,y=0}.
\end{align}
Note that to reproduce the orbit shift in the first formula, the constant term
is not equal to the value of the kick at the bunch center
$k_1(1-e^{-\Delta x^2/2\sigma_2^2})/\Delta x$. It contains
$\sigma_1^2+ \sigma_2^2$ instead of $\sigma_2$ in the exponent.  Therefore,
\cref{eq:111} is not a linear expansion of $\Delta p_{1,x}$.

The linear kick significantly simplifies the analysis. As known from the
accelerator physics, in this case the Gaussian bunch remains Gaussian.  The
offset term is equivalent to the dipole magnet. It shifts the Gaussian {\it
  center} according to \cref{eq:104}. The linear terms
$u\cdot\left.\partial\Delta p_{1,u}/\partial u\right|_{x=\Delta x,y=0}$
represent the quadrupole magnets usually used to focus or defocus the
beams~\cite{Chao:1490001}. They modify the Gaussian {\it widths} according to
the formula
\begin{equation}
\label{eq:112}
\sigma_u' = \sigma_u\sqrt{1 + \frac{\beta_1}{2\tan(2\pi Q_u)}
  \left.\frac{\partial\Delta p_{1,u}}{\partial u}\right|_{x=\Delta x,y=0}}\ \ .
\end{equation}

Another simplification of \cref{eq:111} is that the $x$ and $y$ kicks are
independent, so the beam--beam $x$-$y$ coupling is neglected. Multiple
interaction points also decouple and the analysis can be limited to the point
where the scan is performed. Indeed, other interactions only modify bunch
Gaussian widths and centers, but in any case, they are considered as free
parameters in van der Meer analyses. Since the transverse positions of the
beams at not scanned points are kept constant, the beam--beam modifications
are also constant and do not bias van der Meer calibration. Although the
beam--beam kick at multiple points have been discussed and simulated
in~\cite{madx12}, where the old model was introduced, in practice this was
never used.

For the Gaussian bunches, the luminosity can be calculated from the known
centers and widths.  In the old model the luminosity modification due to the
orbit shift, induced by the equivalent dipole magnet, was calculated
analytically. The contribution from the quadrupole magnet was obtained using
LHC MAD-X accelerator simulation for the beam parameters from
\cref{tbl1}. They were tabulated and then extrapolated analytically to other
bunch settings. Two contributions were summed.

\begin{figure}[htbp]
  \begin{center}
  \includegraphics[width=0.49\textwidth,angle=0]{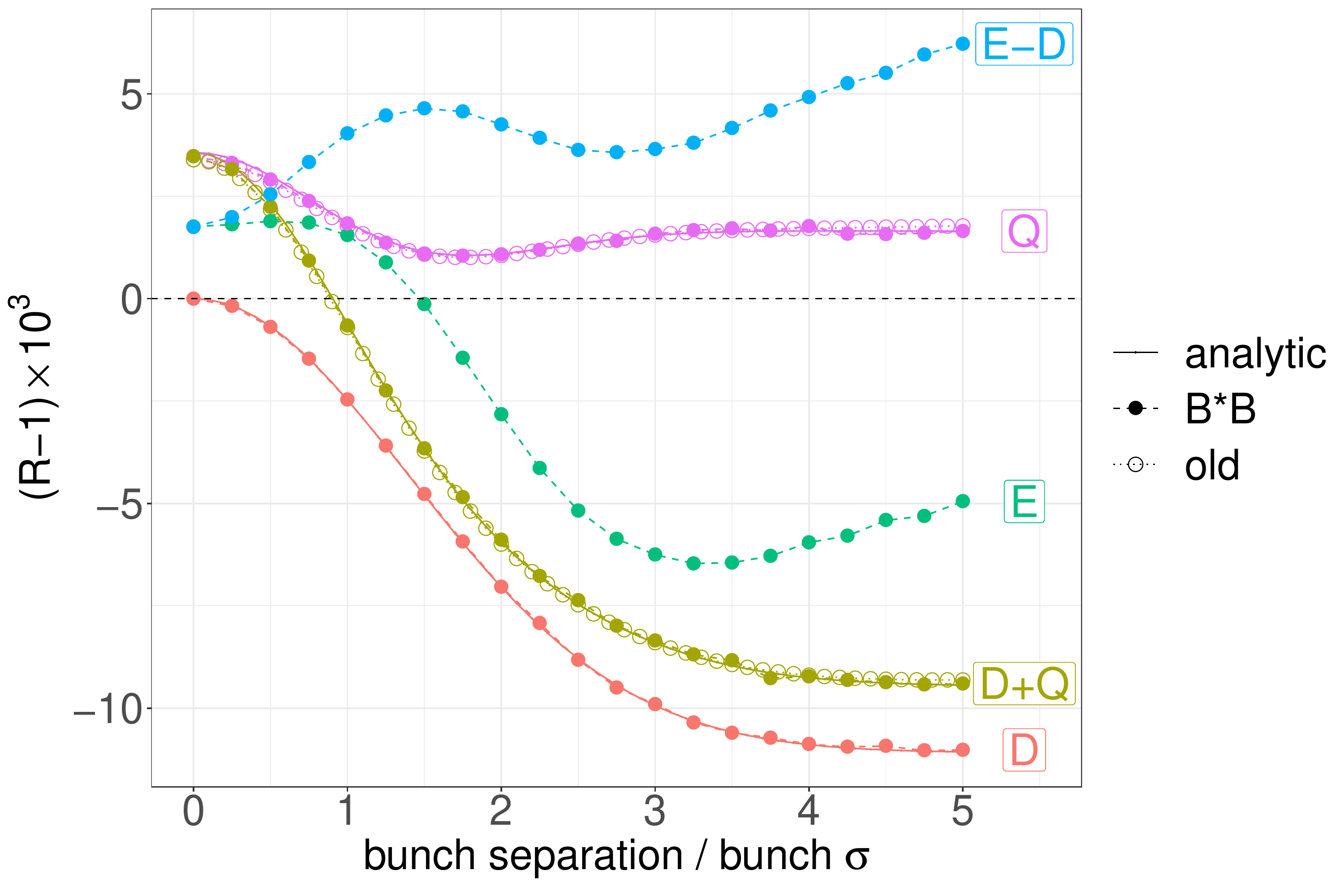}
  \caption{$(R-1)\times10^3$ correction for the bunches with the parameters
    from \cref{tbl1} colliding at one interaction point versus their
    separation in the $x$-scan expressed in the bunch widths. The solid points
    are the results of the B*B simulation when in the recurrence relation
    \cref{eq:81} the exact formula (E), dipole constant dependent on the
    separation (D) or their difference (E-D) are used. The open circles show
    the tabulated values from the old simulation with the quadrupole kick (Q)
    and their sum with the values from the dipole analytic formula (D+Q). The
    curves are analytic predictions for the dipole (D), quadrupole (Q) kicks
    and their sum (D+Q).}
  \label{fig10}
\end{center}
\end{figure}

The results of the old quadrupole simulation for \cref{tbl1} settings are
shown in \cref{fig10} by the open circles. The curves denoted by ``D'', ``Q''
and ``D+Q'' correspond to the dipole, quadrupole contributions and their sum,
respectively.  As in other plots, the luminosity bias $R-1$ from \cref{eq:90}
due to the perturbation of only {\it one bunch} is shown. The {\it analytic}
corrections due to the dipole orbit shifts \cref{eq:104}, the quadrupole bunch
width changes \cref{eq:112} or both are shown by the corresponding solid
lines.  One can see that the quadrupole MAD-X simulation is well described
analytically and \cref{eq:112} could be used instead of the tabulated values
in practice.

To compare these known results with B*B, its exact kick formula was replaced by
\cref{eq:111}, by only its dipole or quadrupole parts. The simulated values
$R-1$ are shown in \cref{fig10} by the solid points.  They are also in good
agreement with the analytic predictions ``D+Q'', ``D''  and ``Q'', respectively.

The B*B results with the {\it exact} kick are shown by the solid circles
connected by the dashed line ``E''. They were already shown in \cref{fig5}.
These results significantly differ from the old ``D+Q'' simulation.  Both
``E'' and ``D+Q'' contain the dipole contribution. To compare with ``Q''
alone, the B*B simulation was performed with the dipole constant
subtracted from the exact kick in the recurrence relation.  The result is
shown as ``E-D'' curve. It differs significantly from ``Q'' and to a larger
extent compensates the luminosity reduction by ``D''.  As expected, the sum of
``D'' and ``E-D'' is in agreement with ``E''.

\begin{figure}[htbp]
  \begin{center}
  \includegraphics[width=0.49\textwidth,angle=0]{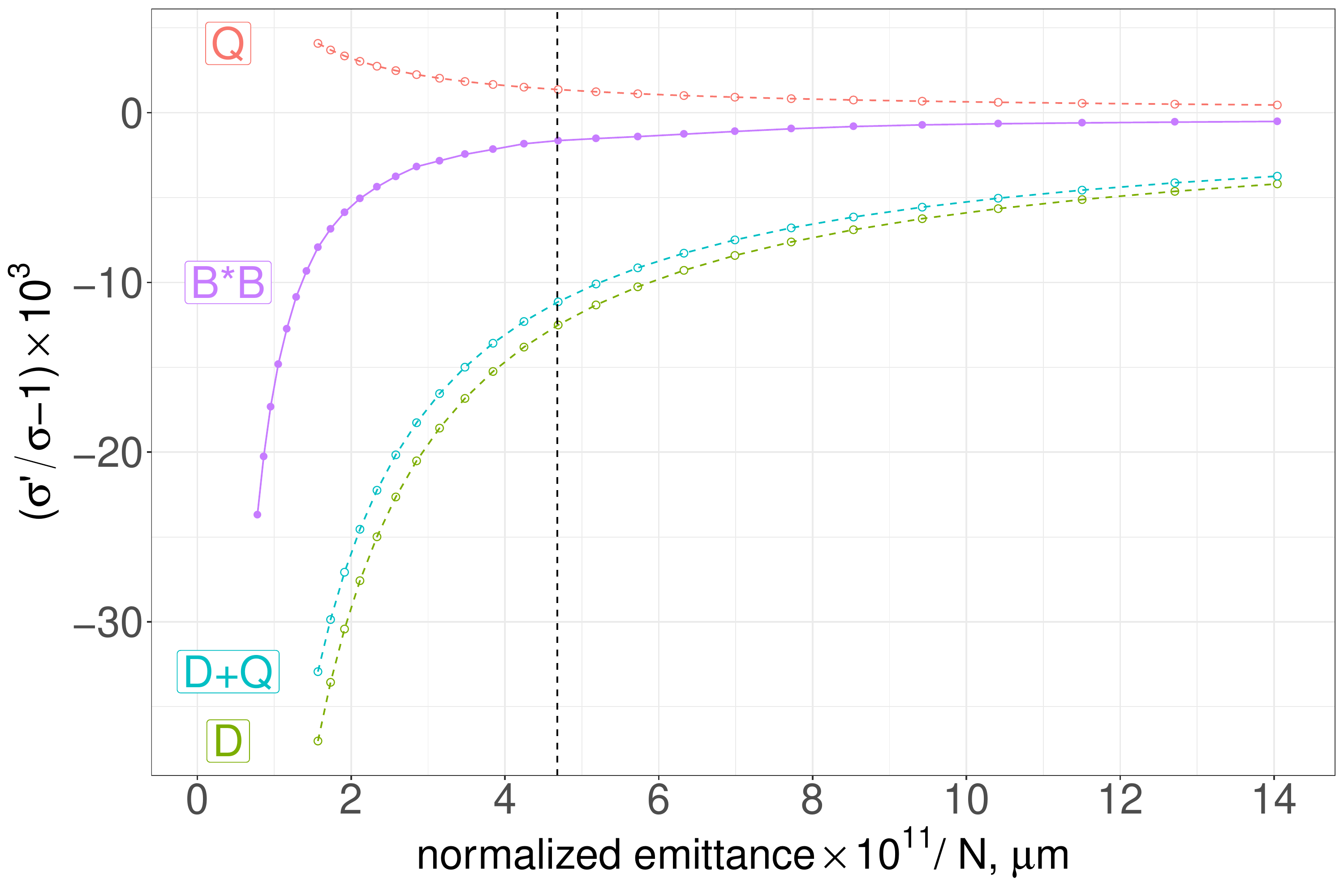}
  \caption{The beam--beam cross-section bias $(\sigma'/\sigma-1)\times10^3$
    determined from \cref{eq:fact} versus $\tilde\epsilon /N\cdot10^{11}$,
    where $\tilde\epsilon$ is the normalized emittance
    $\tilde\epsilon =\epsilon/ (p/m_p)$, identical for both bunches, $m_p$ is
    the proton mass and $N=N_1=N_2$ is the number of protons. The bias is
    calculated for the single interaction point using B*B, analytic dipole
    (D), quadrupole (Q) approximations or their sum (D+Q). The tunes are
    $Q_x=0.31$, $Q_y=0.32$ and it is assumed that
    $\beta_x=\beta_y=\beta$. Under these conditions, $\sigma'/\sigma$ depends
    only on the combination $\tilde\epsilon/ N$ but not on $\epsilon$, $p$,
    $N$ or $\beta$ individually. The vertical dashed line shows the value
    $\tilde\epsilon /N\cdot10^{11}$ for the bunch parameters from
    \cref{tbl1}.}
  \label{fig11}
\end{center}
\end{figure}

The knowledge of the $R$-correction allows determining the bias induced on
the reference cross-section $\sigma$ eg. when using \cref{eq:fact} for the
Gaussian bunches in one-dimensional $x$, $y$ van der Meer scans.  The
integrals in \cref{eq:fact} should be taken over the unperturbed values
\begin{equation}
\label{eq:114}
\mu_{sp}\propto \frac{e^{-\Delta
x^2/2(\sigma_1^2+\sigma_2^2)}}{2\pi\sigma_1\sigma_2}
\end{equation}
modulated by $R^2$ from \cref{fig10} or the corresponding figure for the
$y$-scan.  $R$ should be squared to take into account the perturbation of the
second bunch. The luminosity corrections for the $x$- and $y$-scans are
slightly different because of the difference in the fractional parts of the
tunes $Q_x$, $Q_y$, $0.31\ne0.32$. The resulting biases of van der Meer
cross-section $\sigma'/\sigma-1$ are shown in \cref{fig11} separately for B*B,
dipole (D), quadrupole (Q) approximations and their sum (D+Q).  For the proton
bunches colliding at one point, the correction depends only on the specific
normalized emittance $\tilde\epsilon/N$ assuming identical bunches with
$N=N_1=N_2$ and $\beta_1=\beta_2$. Therefore, this variable is chosen as the
horizontal axis. As shown by the dashed line, the difference between B*B and
the old model for \cref{tbl1} settings is 0.96\%. The old linear kick
approximation was too simple to describe accurately the beam--beam luminosity
bias.


\section{Conclusions}
The main tool for the absolute luminosity calibration at LHC is van der Meer
scan. Its systematics dominates the overall luminosity uncertainty, which, in
turn, gives one of the main contributions to the uncertainty of the accurate
cross-section measurements, for example, in the electroweak sector.

Various details of van der Meer method are discussed in the paper. Currently,
the main sources of systematic uncertainties are the beam orbit drifts, $x$-$y$
non-factorizability and the bunch deformations induced by the beam--beam
electromagnetic interaction.  The first two can be significantly reduced by
the accurate monitoring of the beam positions and the two-dimensional scans,
respectively. The formalism of one- and two-dimensional scans is presented in
the paper in detail.  Other sources of systematics include the measurements of
the bunch populations, the length scale calibration, luminometer detector
effects, other bunch shape changes during the scan, eg. after particle losses,
and various unknown factors contributing to the scan-to-scan
non-reproducibility.  The alternative beam--gas and beam--beam imaging
calibration methods are also briefly mentioned.

The beam--beam bias is the main subject of the paper. The derivation of the
beam--beam kick is presented in \Cref{sec:beambeam} from the first principles
together with the discussion of various approximations and the induced
errors. It is shown that under the assumption of the constant and opposite
velocities of the bunch particles, the calculation of the beam--beam
electromagnetic force reduces to the simple electrostatics between the charges
in the transverse plane. In particular, this allows deriving the average kick
formulas \cref{eq:34,eq:35} for the bunches of arbitrary shapes.

In the last section, we present the B*B simulation for calculating the
beam--beam luminosity corrections. Contrary to the previous model with the
linear kick used at LHC in 2012--2019, it is based on the exact nonlinear
electrostatic force between the point and the Gaussian charge density, either
round \cref{eq:2} or elliptical \cref{eq:BassettiErskine}.  The perturbed
particle trajectories are followed in the accelerator assuming their ideal
transverse betatron motion with the known phase advances. The perturbed
luminosity is calculated at the interaction points with maximally focused
beams, where the derivative of the $\beta$-function is zero, and the
elliptical phase-space trajectories of the betatron motion become circular.

The luminosity corrections due to the perturbations of two bunches are
calculated separately and then summed. It is assumed that the bunch creating
the field is not disturbed by the beam--beam interaction.  Therefore, the
coherent oscillation modes of the two beams are neglected.

The B*B simulation allows to correct the beam--beam luminosity bias in van der
Meer scan point-by-point and to remove it together with the associated $x$-$y$
non-factorizability. The bunch shapes may be approximated by an arbitrary sum
of Gaussians.  The electrostatic field is pre-calculated and then the
interpolations are used to save CPU time.  An arbitrary number of interaction
points is allowed. The simulation of different particles is parallelized
in the processors with multiple cores.  The B*B code is written in C++. It can
be used as a standalone application or as a library available in four computer
languages: C, C++, python and R.

In \cref{sec:inv} it is shown that, for example, the first bunch phase-space
trajectories drawn in the complex planes
$\sqrt{\tilde\epsilon_{1x,1y}/Z_1Z_2N_2}\cdot e^{i\phi_{1x,1y}}$ are
determined only by the initial distribution of
$\tilde\epsilon_{1x,1y}/Z_1Z_2N_2$ ratios, by the phase advances and
$\beta_{1x}/\beta_{1y}$ ratios, if the normalized emittances
$\tilde\epsilon_{1x,1y}=\epsilon_{1x,1y}/(p_1/m_1)$ are conserved. They do not
depend on the individual values of the beta-functions $\beta_{1x,1y}$, the
emittances $\epsilon_{1x,1y}$, the momentum $p_1$, the proton numbers
$Z_{1,2}$ or the number of particles in the opposite bunch $N_2$. This leads
to the corresponding invariance of the luminosity correction ratios.

For the bunch parameters from \cref{tbl1}, ie. for the reference values of the
old model with the linear kick, the B*B simulation predicts 0.96\% less van
der Meer cross-section correction than the old model.  As one can see from
\cref{fig11}, the latter significantly overcorrects the bias. This needs to be
propagated to all LHC cross-sections after 2012 taking into account the bunch
parameters in the corresponding van der Meer calibrations. For other Run 2
proton--proton van der Meer scans the discrepancies are in the range
$0.8-1.4$\%. The B*B predictions are going to be used at LHC in the future
calibration analyses, for example, they already appear in~\cite{atlas20}.
  
\begin{acknowledgements}
  The author wishes to thank the members of LHC Luminosity Calibration and
  Monitoring Working Group 
  and, in particular, Guido Sterbini for calculating the phase advances in
  \cref{tbl2} using LHC MAD-X software in July 2019 and Tatiana Pieloni and
  Claudia Tambasco for confirming \cref{fig4} results using COMBI simulation
  in August 2019.
\end{acknowledgements}


\interlinepenalty=10000

\end{document}